\documentclass[twocolumn]{biophys-new}
\usepackage[utf8]{inputenc}
\usepackage{graphicx}
\usepackage{enumerate}
\usepackage[colorlinks,allcolors=cyan!70!black]{hyperref}
\usepackage{multirow}
\usepackage{makecell}
\usepackage{siunitx}
\usepackage{hyperref}

\graphicspath{{Figures/}}

\title{Path sampling challenges in large biomolecular systems: RETIS and REPPTIS for ABL-imatinib kinetics} %99 characters 
\runningtitle{Path sampling in ABL-imatinib kinetics} 
\author[1]{Wouter Vervust}
\author[2]{Daniel T. Zhang}
\author[3]{Enrico Riccardi}
\author[4]{Titus S. van Erp}
\author[1,*]{An Ghysels}
\runningauthor{Vervust, Zhang, Riccardi, van Erp and Ghysels} %% For page header

\affil[1]{IBiTech - BioMMedA research group, Ghent University, Corneel Heymanslaan 10, entrance 97, 9000 Gent, Belgium}
\affil[2]{Research Institute for Interdisciplinary Science, Okayama University, Okayama 700-8530, Japan}
\affil[3]{Department of Energy Resources, University
of Stavanger, Stavanger, Norway}
\affil[4]{Department of Chemistry, Norwegian University of Science and Technology, NO-7491 Trondheim, Norway}

\corrauthor[*]{an.ghysels@ugent.be}

\papertype{Article}
\begin{document}

\begin{frontmatter}

\begin{abstract}

Predicting the kinetics of drug-protein interactions is crucial for understanding drug efficacy, particularly in personalized medicine, where protein mutations can significantly alter drug residence times. 
This study applies Replica Exchange Transition Interface Sampling (RETIS) and its Partial Path variant (REPPTIS) to investigate the dissociation kinetics of imatinib from Abelson nonreceptor tyrosine kinase (ABL) and mutants relevant to chronic myeloid leukemia therapy.
These path-sampling methods offer a bias-free alternative to conventional approaches requiring qualitative predefined reaction coordinates. Nevertheless, the complex free-energy landscape of ABL-imatinib dissociation presents significant challenges. 
Multiple metastable states and orthogonal barriers lead to parallel unbinding pathways, complicating convergence in TIS-based methods. 
Despite employing computational efficiency strategies such as asynchronous replica exchange, full convergence remained elusive.
This work provides a critical assessment of path sampling in high-dimensional biological systems, discussing the need for enhanced initialization strategies, advanced Monte Carlo path generation moves, and machine learning-derived reaction coordinates to improve kinetic predictions of drug dissociation with minimal prior knowledge. 

%TODO incorporate these citations
%https://pubs.acs.org/doi/full/10.1021/acs.jpclett.4c02332
%https://pubs.acs.org/doi/full/10.1021/acs.jctc.4c01108

\end{abstract}

\begin{sigstatement}
%120 words allowed. This is 110 words
%This study investigates the dissociation kinetics of imatinib from ABL kinase using RETIS and REPPTIS, marking a significant application of these path sampling methods to a complex biomolecular system. Despite challenges posed by rugged energy landscapes and metastable states, substantial advancements were made in optimizing the calculation of complex order parameters, greatly improving computational efficiency. The findings highlight critical limitations of one-dimensional reaction coordinates and underscore the need for higher-dimensional approaches and better initialization strategies, paving the way for methodological innovations in biomolecular kinetics and computational drug design.
%%%
%%%
% more focus on ABL, less on computational efficiency:
This study examines the dissociation kinetics of imatinib from ABL kinase using RETIS and REPPTIS, assessing the feasibility of path sampling methods for drug unbinding in a clinically relevant system. ABL mutations drive imatinib resistance in chronic myeloid leukemia, underscoring the need for computational tools that predict mutation-induced kinetic effects. However, the complex energy landscape, metastable states, and parallel unbinding pathways posed significant challenges, revealing limitations of one-dimensional reaction coordinates. This work highlights the need for improved initialization strategies and higher-dimensional approaches to enhance the predictive power of path sampling in biomolecular kinetics, providing insights that guide future advancements in computational drug design and personalized medicine.
\end{sigstatement}
\end{frontmatter}

%----------------------------------
\section*{Introduction}
%----------------------------------

Traditionally, equilibrium thermodynamic properties such as
binding affinity
(dissociation constant $K_d$) and IC$_{50}$ (concentration causing 
50\% target inhibition)
have dominated as primary predictors of drug 
efficacy~\cite{lipinski2012experimental,jorgensen2004many,
claveria2017look,mobley2017predicting}.
However, drug kinetics, 
particularly drug residence time, 
has been increasingly recognized for its crucial role in 
pharmacodynamics and better correlation 
with \textit{in vivo} 
efficacy~\cite{copeland2006drug,tummino2008residence,copeland2010dynamics,
lu2010drug,copeland2016drug}. 

Personalized medicine is particularly relevant for kinase inhibitors, 
where protein mutations can drastically alter drug residence times, 
affecting therapeutic effectiveness~\cite{wang2011genomics,
relling2015pharmacogenomics,sneha2016molecular}. 
A well-known example is Abelson nonreceptor tyrosine kinase (ABL), 
which plays a significant role in chronic myeloid leukemia (CML)
through its fusion oncogene BCR-ABL~\cite{georgoulia2019catalytic,o2012pushing}.
The first-generation tyrosine kinase inhibitor (TKI) imatinib 
(Gleevec\textsuperscript{\tiny\textregistered}, also known as STI571) was
developed to competitively bind to the ATP-binding site. Imatinib demonstrated remarkable efficacy in the treatment of CML and is recognized as a
landmark achievement in targeted drug
design~\cite{georgoulia2019catalytic,o2012pushing,stegmeier2010targeted}.
However, mutations in the ABL kinase domain led to resistance, necessitating the development of second- and third-generation 
TKIs~\cite{o2012pushing,reddy2012ins}. 
This iterative arms race between targeted therapy and resistance mutations underscores the necessity of predictive tools that assess how mutations drive changes in drug kinetics.

Given the costly and time-consuming nature of experimental residence time
assays, computational methods have emerged as a promising alternative in the early stages of the drug-design 
pipeline ~\cite{shoichet2004virtual,copeland2006drug,okimoto2009high,
jorgensen2009efficient,lin2020review}.
Molecular dynamics (MD) simulations can provide valuable insight into 
the mechanisms of drug unbinding and even provide a timescale on drug dissociation. 
%An example is the unbinding rate of a drug molecule from its target protein, which is a measure for the residence time and thus efficacy of the drug. 
The unbinding mechanism is, however, often a rare event, where dissociation is preceded by a long waiting time -- in terms of computer simulation wall time. Moreover, multiple occurrences need to be observed to obtain reasonable statistics. Even with the current high performance computing infrastructure, this prevents the application of standard MD simulations to assess and predict timescales of biological processes that occur less frequently than several times per 10\,$\mu$s.
For instance, predicting a drug residence time of 1\,ms lies beyond what is currently feasible.

To overcome this, path sampling methodologies have been developed to 
efficiently explore rare events by focusing computational effort on 
transition events rather than waiting times~\cite{dellago1998transition,bolhuis2002transition,bolhuis2021transition}. 
Several methods have been developed based on path ensembles~\cite{bolhuis_transition_2021}, including transition interface sampling (TIS)~\cite{van2003novel}, which gives exact kinetics~\cite{van2012dynamical}.
%. and forward flux sampling (FFS) ~\cite{faradjian2004computing,allen2005sampling,allen2009forward}. 
%Both FFS and TIS are exact methods but differ fundamentally in their algorithmic approach. FFS can handle nonequilibrium driven systems, unlike TIS, but requires stochastic dynamics, whereas TIS also works for deterministic dynamics. Moreover, FFS is more susceptible than TIS to sampling unrepresentative transition paths~\cite{van2012dynamical}.  
Milestoning~\cite{faradjian2004computing,ojha2024advances} 
and the partial path variant of TIS (PPTIS)\cite{moroni2004rate} introduce a Markovian approximation, statistically combining short trajectory segments to reconstruct the full transition efficiently, especially for rare and long transitions. 
The efficiency of the exact TIS methods can be enhanced by incorporating swaps between ensembles, as implemented in replica exchange TIS (RETIS)~\cite{van2007reaction,bolhuis2008rare,cabriolu2017foundations}, in its partial path counterpart called replica exchange PPTIS (REPPTIS)~\cite{vervust2023path}, and in the 
recently developed infinite swap and asynchronous version called $\infty$RETIS \cite{zhang2024highly}.
%transition interface sampling (TIS)~\cite{van2003novel}, and replica exchange TIS (RETIS) where replica exchange between different ensembles increases sampling efficiency~\cite{van2007reaction,bolhuis2008rare,cabriolu2017foundations},partial path TIS (PPTIS) where paths are cut short~\cite{moroni2004rate},replica exchange partial path TIS (REPPTIS)~\cite{vervust2023path},and milestoning with memory loss at the interfaces, also known as milestones~\cite{faradjian2004computing}.
%Among these methods, (RE)TIS is in principle giving the exact kinetis, and it has
(RE)TIS methods have
been successfully applied to study chemical reactions~\cite{Mahmoud_silic, moqadamlocal2018, daub2020path}, thin film breakage~\cite{aaroen2022thin,zhang2023enhanced}, solid-solid phase transitions in crystal structures~\cite{lervik2022role},
protein-DNA interaction~\cite{riccardi2019predicting}
and permeation of oxygen through membranes~\cite{riccardi2020permeation,Ghysels2021exactnonMarkov}.
However, their applicability to large biological systems, such as kinase-drug dissociation, remains largely untested. 

In this work,
we explore the application of RETIS, REPPTIS, and $\infty$RETIS to imatinib dissociation from wild-type and mutant ABL kinase domains, assessing the 
feasibility of these methods for studying mutation-driven effects on drug kinetics. 
Imatinib dissociation from ABL is found to be characterized by the presence of metastable states and multiple parallel dissociation pathways, which introduce significant challenges for the current 
TIS-based sampling methodologies, including sampling bottlenecks, slow convergence, and difficulties in defining optimal reaction coordinates. 
Our paper highlights these key computational challenges and offers an outlook on 
necessary methodological developments to improve the predictive capabilities of RE(PP)TIS for drug dissociation kinetics. 

The remainder of this work is structured as follows. In the \hyperref[sec:methods]{Methods}, the ABL-imatinib complex is briefly described, followed by a description of the (RE)(PP)TIS methodologies. The equilibrium simulations, the order parameter definitions, and the steered simulations that are required to initialize the path sampling simulations are then discussed. The section ends with optimizations that were implemented to speed up the path sampling simulations.
The \hyperref[sec:results]{Results} show how metastable states and parallel distinct reaction channels characterize imatinib dissociation, and it is discussed how these two factors limit the current RE(PP)TIS methodology in the \hyperref[sec:discussion]{Discussion}. The \hyperref[sec:conclusion]{Future Perspectives and Conclusion} then give an outlook on how to tackle these challenges. 

\section*{Methods}\label{sec:methods}
%-------------------------

\subsection*{The ABL-imatinib complex}
%-------------------------

\begin{figure}[htb]
    %\centering
    \includegraphics[width=8cm]{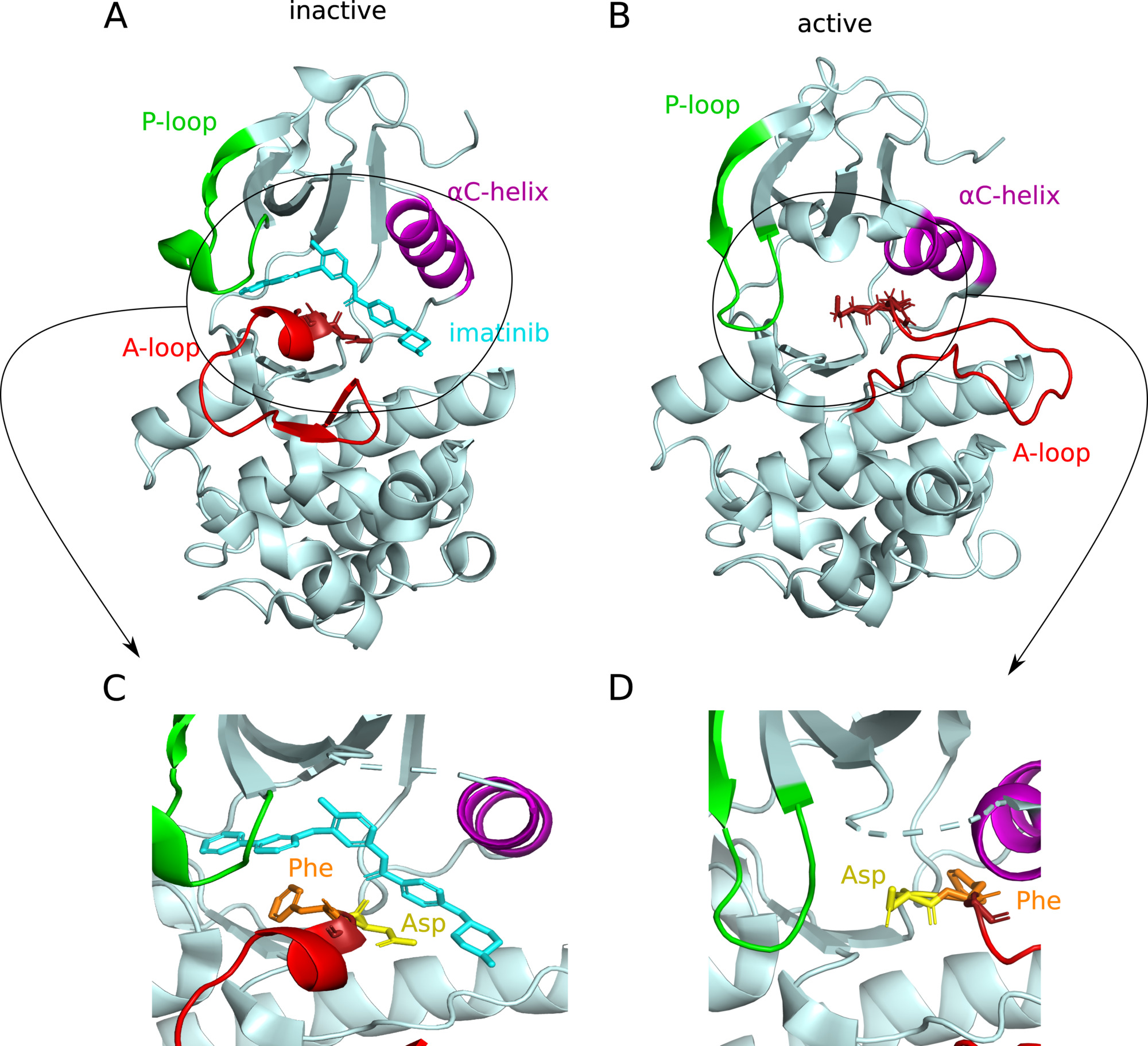}
    \caption{Inactive (\textbf{A}, PDB 6NPV~\cite{simpson2019identification}) 
    and active (\textbf{B}, PDB 6XR6~\cite{xie2020conformational}) 
    conformations of the ABL kinase domain. \textbf{C}: Close up of the binding 
    pocket in the inactive state. Imatinib binds to the 
    inactive state, where the aspartate residue points outwards of the binding
    pocket. \textbf{D}: Close up of the binding pocket in the active state. 
    In the active state, the aspartate residue points
    inwards of the binding pocket. In \textbf{A} and \textbf{B}, the dashed cyan 
    lines 
    represent missing residues of the 6NPV structure. 
    In \textbf{D}, the dashed cyan lines represent removed residues for better 
    visualization of the binding pocket. Figure created with
    PyMOL~\cite{delano2002pymol}.}
    \label{fig:ABL:open_closed}
\end{figure}

The ABL kinase domain is shown in Fig.~\ref{fig:ABL:open_closed}, 
consisting of the N-terminal 
and C-terminal lobes typical for protein kinases. Three important regions 
are highlighted: the activation loop (A-loop, red), the ATP-binding loop
(P-loop, green), and the $\alpha$C-helix (magenta). The N-lobe consists 
of a five stranded $\beta$-sheet and the $\alpha$C-helix, and also 
contains the P-loop. The 
C-lobe mainly consists of helical structures and contains the substrate 
binding site~\cite{panjarian2013structure}. The active site (ATP-binding site) 
is located in the cleft between the N- and C-lobes. 

Located at the N-terminal side of the A-loop is the 
aspartate-phenylalanine-glycine (DFG) motif. The active state 
is characterized by the aspartate pointing inwards towards the ATP-binding 
site (DFG-in, Fig.~\ref{fig:ABL:open_closed}D). 
Transitioning to the inactive state is accompanied by the 
DFG motif flipping \qty{180}{\degree}, where the aspartate residue moves 
outwards and the 
phenylalanine moves inwards 
(DFG-out, Fig.~\ref{fig:ABL:open_closed}C)~\cite{panjarian2013structure}. 
In the active state, the $\alpha$C-helix is turned away from the hinge 
region, and the A-loop is in an open (extended) conformation.
In the inactive state, the $\alpha$C-helix is turned towards the 
hinge region, and the A-loop can block substrate binding to the C-lobe.

The ABL kinase domain catalyzes the transfer of the $\gamma$-phosphate 
from adenosine triphosphate (ATP) to tyrosine residues in substrate
proteins~\cite{panjarian2013structure}. 
Patients with the fusion oncogene BCR-ABL
suffer from chronic myeloid leukemia (CML), where the normally autoinhibited 
kinase domain becomes constitutively active~\cite{georgoulia2019catalytic,o2012pushing}.
Imatinib inhibits this activity by competitively binding to the ATP-binding site, thereby preventing ATP from engaging with the kinase. 
However, ABL mutations can confer resistance by destabilizing imatinib 
binding or favoring conformations that reduce drug
affinity~\cite{shah2002multiple,shah2007sequential}.
To computationally assess how such mutations influence imatinib residence time, REPPTIS and 
$\infty$RETIS are used to investigate imatinib dissociation in both the wild-type (WT) ABL kinase and seven commonly encountered mutated variants for which experimental dissociation constants are available (Table~\ref{tab:ABL:mutations}).

%In this study, the dissociation of imatinib from the wild-type (WT) ABL kinase 
%domain is 
%investigated using the REPPTIS and $\infty$RETIS methods.
%Furthermore, $7$ mutated variants 
%of ABL are studied (Table~\ref{tab:ABL:mutations}). 
%To run these simulations, 
%initial paths for the ensembles are required, which are obtained by first 
%running equilibrium simulations followed by steered MD simulations.

\begin{table}[htb]
    \centering
    \begin{tabular}{llll}
    \hline
    Variant & H$_2$O   & Na$^+$ & Cl$^-$  \\
    \hline
    WT       & 17355 & 61 & 53  \\
    Q252H    & 17359 & 61 & 53  \\
    Y253F    & 17360 & 61 & 53  \\
    E255V    & 17362 & 60 & 53  \\
    T315I    & 18858 & 62 & 56  \\
    M351T    & 17358 & 61 & 53  \\
    F359V    & 17364 & 61 & 53  \\    
    H396P    & 17358 & 61 & 53 \\
    \hline
    \end{tabular}
    \caption{The ABL-imatinib variants studied in this work. WT denotes the wild-type ABL kinase. The columns specify the number of H$_2$O molecules, Na$^+$ ions, and Cl$^-$ ions used in the MD simulation boxes. Variants were selected based on whether PBD structures and experimental dissociation constants ($K_D$) with imatinib were available, after which seven were selected at different residue index positions.}
    \label{tab:ABL:mutations}
\end{table}

\subsection*{RETIS and REPPTIS theory}
%-------------------------
A brief overview of the TIS and PPTIS methodologies, including the (infinite) replica exchange formalism, is now given. Additional information is provided in Text~\hyperref[sec:si]{S}1, while in-depth discussions can be found in Refs.~\cite{cabriolu2017foundations,moroni2004rate,vervust2023path,van2023far,zhang2024highly}.
In both TIS and PPTIS, an order parameter $\lambda$ is defined to track the progression of the reaction, along which
$n+1$ non-intersecting hypersurfaces (interfaces) $\{x|\lambda(x)=\lambda_i\},  \forall i\in\{0,\dots,n\}$ are defined (Fig.~\hyperref[sec:si]{S}1), where $x$ denotes the phase point of the system. The interfaces are monotonically increasing ($\lambda_i < \lambda_{i+1}$), where the region left of $\lambda_0\equiv\lambda_A$ defines state $A$ ($A$ is $\{x\,|\,\lambda(x)<\lambda_A\}$) 
and the region right of $\lambda_{n}\equiv\lambda_B$ defines state $B$ ($B$ is $\{x\,|\,\lambda(x)>\lambda_B\}$).

To each interface $\lambda_i$, a path ensemble $[i^+]$ (TIS, Fig.~\hyperref[sec:si]{S}1B) or $[i^\pm]$ (PPTIS, Fig.~\hyperref[sec:si]{S}1C) is associated. For TIS, the $[i^+]$ ensemble consists of all the paths that start at $\lambda_A$, then cross $\lambda_i$ before recrossing $\lambda_A$, and end in either $\lambda_B$ or $\lambda_A$. For PPTIS, the paths of $[i^\pm]$ are also required to cross $\lambda_i$, but the starting and ending conditions are set at the neighboring interfaces $\lambda_{i-1}$ and $\lambda_{i+1}$, meaning that the paths are cut short. For both TIS and PPTIS, the ensemble associated to $\lambda_i$ allows the calculation of a (conditional) local crossing probability, which is the chance that a path that has just crossed $\lambda_i$ will cross the next interface $\lambda_{i+1}$ rather than going back to $\lambda_A$ (TIS) or $\lambda_{i-1}$ (PPTIS). 
For TIS, the crossing behavior around $\lambda_i$ is thus analyzed using paths that started with escaping $A$, meaning that all `path memory' is incorporated, resulting in an essentially exact kinetic reconstruction. 
This reconstruction is done by viewing the rate constant $k_{AB}$ as the product of a positive flux $f_A$ (how many times does the system escape state $A$, per unit of time) and a global crossing probability $P_A(\lambda_B|\lambda_A)$ (how likely is an escape from state $A$ to result in reaching state $B$, rather than returning to state $A$)
\begin{equation}
k_{AB} = f_A\times P_A(\lambda_B|\lambda_A).
\label{eq:rate}
\end{equation}
The flux $f_A$ can be easily computed from a straightforward MD simulation, as in TIS and PPTIS, or, as in RETIS, from the average path lengths in the ensembles $[0^+]$ and the RETIS-specific path ensemble $[0^-]$, which explores the left side of $\lambda_A$. The global crossing probability, however, is typically too small to be computed directly. Instead, it is expressed as a product of local conditional crossing probabilities $P_A(\lambda_{i+1}|\lambda_i)$, which represent the probability of reaching $\lambda_{i+1}$ given that $\lambda_i$ has been reached. This probability is equivalent to the fraction of sampled paths in $[i^+]$ that cross $\lambda_{i+1}$. The number and placement of interfaces are chosen such that $P_A(\lambda_{i+1}|\lambda_i)$ are sufficiently large to be estimated accurately with a relatively modest number of trajectories.
In this work, the number of interfaces and their placements for the ABL-imatinib systems
were first estimated from a rough free energy calculation using umbrella sampling (US), as shown in Fig.~\hyperref[sec:si]{S}6.

The catch for TIS lies in the presence of metastable states, where paths can get stuck and commit neither to $A$ nor $B$, resulting in (computationally) infeasible long paths. For PPTIS, the crossing behavior around $\lambda_i$ is analyzed using shorter paths whose memory has been cut (Text~\hyperref[sec:si]{S}1, Fig.~\hyperref[sec:si]{S}1C), resulting in an essentially approximative reconstruction of the kinetics that is, however, faster and more feasible when barrier-crossing trajectories are not only rare but also very long. The total crossing probability in PPTIS is calculated from recursive relations that use as input the local conditional crossing probabilities $p_i^\pm$, $p_i^=$, $p_i^\mp$, and $p_i^\ddagger$. Here, the top plus or minus sign indicates the probability of transitioning from $\lambda_i$ to $\lambda_{i+1}$ or $\lambda_{i-1}$, respectively, while the bottom sign indicates the previous state of the system before crossing $\lambda_i$.

Practically, the ensembles $[i^+]$ (TIS) or $[i^\pm]$ (PPTIS) are sampled using a Monte Carlo procedure. New paths are generated from old ones using shooting moves (or more advanced variants~\cite{grunwald2008precision,gingrich2015preserving,jung2017transition,menzl2016s,borrero2016avoiding,riccardi2017fast,zhang2023enhanced}. A different kind of move is given by replica exchange, where it is attempted to swap paths between different ensembles (Text~\hyperref[sec:si]{S}1, Fig.~\hyperref[sec:si]{S}2), resulting in an increase of ergodic sampling and faster convergence for both RETIS and REPPTIS~\cite{van2007reaction, vervust2023path}. 
The recently released $\infty$RETIS method maximizes the benefits of replica exchange by considering all (infinite) possible swaps after every shooting-like move~\cite{roet2022exchanging,zhang2023enhanced}. This approach is particularly suited for parallel computing, where replica exchange moves can be executed asynchronously. Achieving the infinite swapping limit at each step enhances sampling efficiency, making $\infty$RETIS a highly efficient method for advanced molecular simulations~\cite{roet2022exchanging, zhang2024highly}.

Application of RE(PP)TIS to the ABL-imatinib variants requires the definition of an order parameter and the construction of initial paths to which the MC moves can be applied. The initial paths are constructed using steered MD simulations preceded by equilibrium simulations, which are detailed in the following sections. 

\subsection*{Equilibrium simulations}

The crystal structure of the ABL kinase domain in complex with imatinib was 
obtained from the 6NPV PDB entry~\cite{simpson2019identification}. 
As there 
were no structures available for $6$ of the $7$ mutated ABL variants 
in complex with 
imatinib 
or similar TKIs, CHARMM-GUI~\cite{jo2008charmm} was used to mutate 
the residues based on the 6NPV structure. For the gatekeeper mutation T315I, 
a crystal structure was available in complex with the inhibitor DCC-2036 
(PDB entry 3QRJ
~\cite{chan2011conformational}). 
To create the T315I variant, the 3QRJ structure 
was superimposed on the 6NPV structure, the DCC-2036 ligand was removed,
and the imatinib ligand added, as shown in Fig.~\hyperref[sec:si]{S}4.

The structures were then prepared for MD simulations using the 
GROMACS~\cite{abraham2015gromacs}
software package (version 2021.3). 
The CHARMM36m force field~\cite{huang2017charmm36m} was used for the protein
and solvent, where the TIP3P water
model was used~\cite{jorgensen1983comparison}. The force field parameters for 
imatinib were taken from Ref.~\cite{lin2014computational}, which 
were obtained using QM calculations using the GAAMP software
~\cite{huang2013automated,boulanger2018optimized} and which were 
recently used in another study on the WT ABL-imatinib complex
~\cite{paul2020diversity}. The ligand force field was provided in CHARMM 
format, and it was converted to the GROMACS format using CHARMM-GUI
~\cite{jo2008charmm}. The imatinib parameters define a protonated state, which 
is desirable as at physiological pH the complex with neutral imatinib 
represents less than 0.1\% of the overall
population~\cite{aleksandrov2010molecular}.

The structures were solvated in a rhombic dodecahedron box,
ensuring a minimum distance of 1.5 nm between periodic images. 
The dodecahedral box is only 71\% of the cubic box volume with the same minimum 
distance criterion, resulting in significant computational savings. The minimal 
image distance is chosen to be larger than the typical 1 nm value 
as dissociation is to be studied. 
Na$^+$ and Cl$^-$ ions were added to neutralize the
system and to reach a physiological 
salt concentration of 0.15 M. The water and ion content of the 
simulation boxes is shown in Table~\ref{tab:ABL:mutations}. 
The systems were then energy minimized using steepest descent 
until the maximum force was below 
1000 kJ mol$^{-1}$ nm$^{-1}$. 
The systems were then equilibrated in the NVT and NPT ensemble for 100 ps,
each,  
where both the ligand and protein heavy atoms were restrained. 
The equilibrium production simulations (unrestrained) were run for 50 ns each, 
in the NPT ensemble
at 300 K and 
at 1 bar using the Bussi–Donadio–Parrinello thermostat
~\cite{bussi2007canonical} with a coupling constant of 0.1 ps and the
Parinello-Rahman barostat~\cite{parrinello1981polymorphic} 
with a coupling constant of 2 ps and 
compressibility of 4.5e-5 bar$^{-1}$. 
Hydrogen bonds were constrained using LINCS~\cite{hess1997lincs},
which allowed for a time step of 2 fs.

\begin{figure}[htb!]
    \centering
    \includegraphics[width=0.48\textwidth]{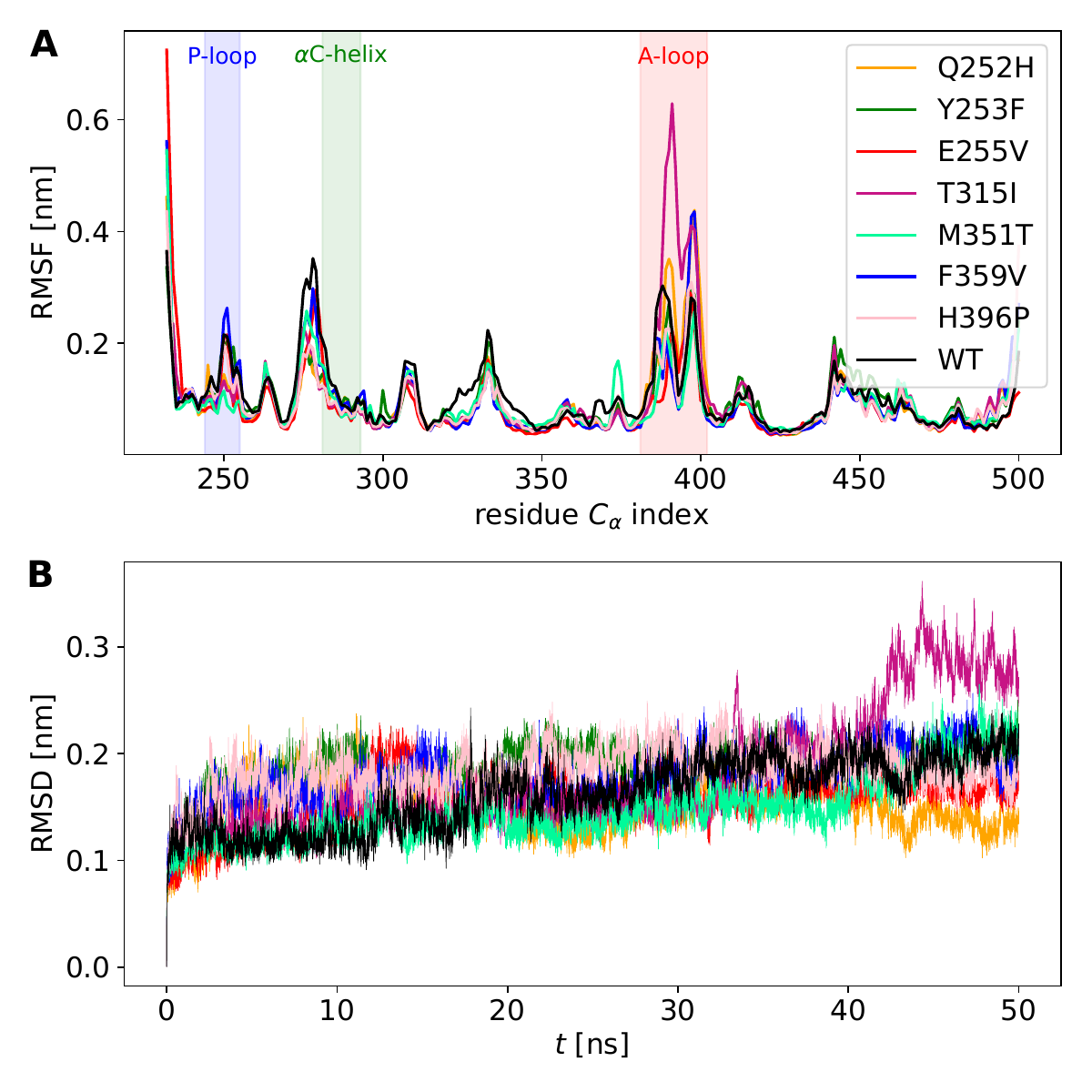}
    \caption{RMSF (\textbf{A}) and RMSD (\textbf{B}) 
    of the C$_\alpha$ atoms in equilibrium MD simulations, 
    for native ABL
    (WT, black line) and the $7$ mutated variants (colored lines). The alignment was performed using the
    rigid C$_\alpha$'s.
    The first 
    frame of each equilibrium run served as reference for the RMSD.
    %(as defined in 
    %Sec.~\ref{sec:ABL:order_parameter}).
    Both figures share the same legend. Residues are numbered according to the
    UniprotKB accession number P00519 (tyrosine-protein kinase ABL1).}
    \label{fig:ABL:RMSF}
\end{figure}

The equilibrium production simulations were checked for validity 
(temperature, energy, pressure) and 
by visualizing the trajectories with the VMD~\cite{humphrey1996vmd} software.
A set of rigid C$_\alpha$'s was determined for each variant based on the fluctuations in the 50 ns equilibrium MD run (Table~\hyperref[sec:si]{S}1).
The root mean square deviation (RMSD) and root mean square fluctuations (RMSF) 
of the protein C$_{\alpha}$'s are shown in Fig.~\ref{fig:ABL:RMSF}, where 
protein alignment was performed using the rigid C$_\alpha$'s only, and where 
the first frame served as reference.

\subsection*{Order parameter definition}\label{sec:ABL:order_parameter}

The order parameter $\lambda(x)$ for a configuration (snapshot, frame) $x$ 
is defined 
as the distance between the current center of mass (COM) of imatinib 
$\mathbf{r}_\text{COM}(x)$ and the 
average bound COM position of imatinib $\mathbf{r}_\text{COM}^{\text{bound}}$, 
\begin{equation}
\lambda(t) = \left\lVert\mathbf{r}_\text{COM}(t) - 
\mathbf{r}_\text{COM}^{\text{bound}}\right\rVert
\label{eq:op}
\end{equation}
where $\mathbf{r}_\text{COM}^{\text{bound}}$ approximates the natural bound position of imatinib, which is defined as follows. 
First, for each variant,
the frames of the 50 ns equilibrium trajectory were 
superposed on the final frame of the trajectory using the rigid C$_\alpha$'s (i.e.\ the reference frame $x_\text{ref}$ was the final frame). Second, the average COM position of imatinib in the superimposed frames was calculated and defined as 
$\mathbf{r}_\text{COM}^{\text{bound}}$. 

\begin{figure}[htb]
    \centering
    \includegraphics[width=0.48\textwidth]{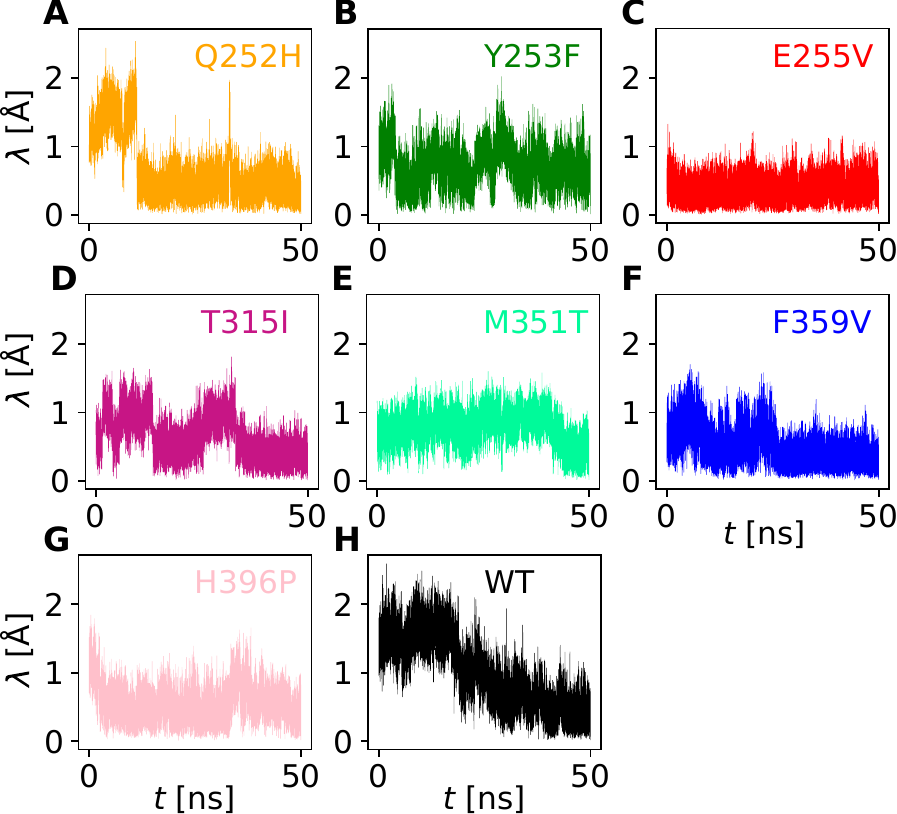}
    \caption{Order parameter $\lambda$ (Eq.~\ref{eq:op}) for equilibrium MD trajectories of the ABL variants.}
    \label{fig:ABL:lambda_trajs}
\end{figure}

The order parameter $\lambda(x(t))$ of a new frame $x(t)$ can now be calculated 
by first superposing this new frame to the reference frame $x_\text{ref}$ using the rigid C$_\alpha$'s,
after which the current COM position of imatinib $\mathbf{r}_\text{COM}(t)$ is calculated,
and its distance 
to $\mathbf{r}_\text{COM}^{\text{bound}}$ is calculated (Eq.~\ref{eq:op}).
As an example, the order parameters are calculated for the equilibrium MD trajectories and shown in 
Fig.~\ref{fig:ABL:lambda_trajs}, where metastable states are seen to be present 
even in the deep binding pocket of ABL. This is particularly visible for the 
Q252H variant (Fig.~\ref{fig:ABL:lambda_trajs}A), where the first 12\,ns of 
the trajectory are spent in a metastable state (Fig.~\hyperref[sec:si]{S}5) that is clearly separated 
from the remaining 38 ns. 
It was chosen not to incorporate the 
initial 12\,ns of the Q252H equilibrium run for the definition of its
average bound COM position 
$\mathbf{r}_\text{COM}^{\text{bound}}$.

\subsection*{Pulling simulations to generate initial path}\label{sec:ABL:steeredMD}

To create an initial path to start the path sampling,
pulling simulations (steered MD) were performed with GROMACS using the
PLUMED~\cite{tribello2014plumed} extension (version 2.7.2). 
Care was taken to ensure that the pulling simulations did not introduce 
large artifacts in the dissociation pathways, which involved manually checking 
the trajectories and several reruns with optimized parameters. 
Artifacts were encountered in two ways. First, it was noticed that pulling 
without restraints on the protein resulted in the C-lobe and N-lobe 
separating (the hinge region completely opening up). Second, it was noticed 
that the pulling force and velocity should be chosen carefully. If these values 
are too high, the system does not have time to relax 
along the orthogonal degrees of freedom,
which can greatly impact the equilibration 
time of REPPTIS.

\begin{figure}[htb!]
    %\centering
    \includegraphics[width=0.48\textwidth]{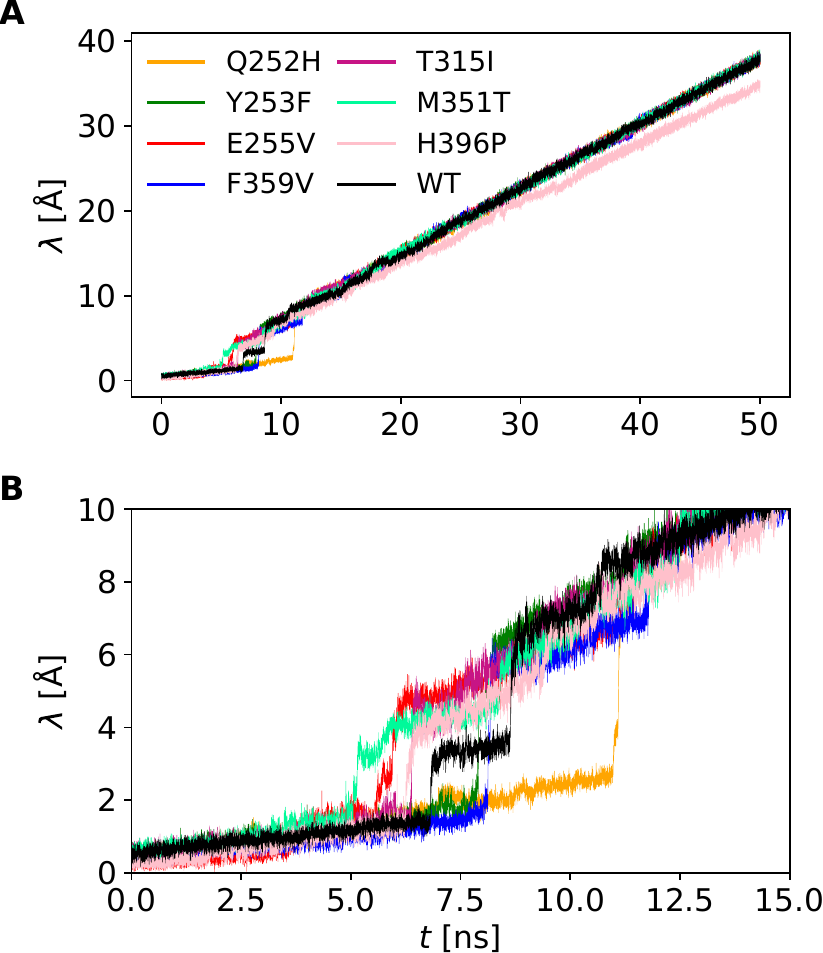}
    \caption{Pulling simulations of 50 ns, for each variant. 
    %\textbf{A}: The potential energy of the systems is void of large artifacts. The T315I variant (purple) has a larger simulation box and hence lower potential energy.
    \textbf{A}: Order parameter $\lambda$. \textbf{B}: Zoom-in on the first 15 ns of dissociation,
    showing `jumping' behavior typical of activated processes. Native ABL 
    (black line) shows a double transition, indicating the presence of 
    a strongly bound metastable state.}
    \label{fig:ABL:pull}
\end{figure}

Pulling trajectories were obtained by restraining the 
C$_\alpha$'s of the rigid residues ($\kappa$ = 5000 kJ mol$^{-1}$nm$^{-2}$)
and by pulling along the order parameter $\lambda$ with a moving restraint 
($\kappa$ = 250 kJ mol$^{-1}$ nm$^{-2}$) aimed to displace imatinib 
38 \AA\ from the bound position in 50 ns (i.e.\ a pulling
velocity of 0.76 \AA\,ns$^{-1}$). 
As the pulling was done slowly, and 
$\lambda$ is defined as the distance to the average bound state, it is expected 
that the pulling force was non-directional. Six of the variants escaped 
via a pathway under the $\alpha$C-helix, while the Q252H and F359V variants 
had imatinib escape via a pathway under the P-loop (Text~\hyperref[sec:si]{S}3, Fig.~\hyperref[sec:si]{S}3). 
%The potential energy of the systems during the pulling simulations is shown in Fig.~\ref{fig:ABL:pull}A, where the perturbation introduced by the pulling force is hardly visible. 
The evolution of the order parameter
$\lambda$ during pulling is shown in Figs.~\ref{fig:ABL:pull}A-B. 
%where the `rareness' (i.e.\ sudden quick jumps in $\lambda$ indicate quick transits over a steep barrier.  % REMARK: discutable, interpretation not that straightforward.
The quick jumps in the dissociation are seen to be conserved in the variants
(Fig.~\ref{fig:ABL:pull}B). 
If the pulling force is too high, this plot would contain straight lines, where dissociation is more likely to sample unlikely pathways. 
Imatinib is strongly bound in the binding pocket of ABL, as suggested by a sudden 
transition (jump in $\lambda$) around the 
5 ns to 10 ns mark, 
depending on the variant. Most of the variants do not simply display a single 
transition, and e.g.\ 
a double transition for WT ABL is clearly visible. This suggests the presence of
metastable states, some of which have imatinib still strongly bound to ABL.
Once imatinib transitions out of 
the deep binding pocket and its metastable states, 
the pulling profile is rather linear. This suggests that, at least for these 
pathways, the ligand was no longer strongly bound to the protein.

%\subsection*{Umbrella sampling simulation for interface placement}

%An essential part to run any TISbased method is to set the number of interfaces and the positions of these interfaces. As a preparation for running the RETIS, REPPTIS, and $\infty$RETIS simulations, the free energy as a function of the order parameter $\lambda$ was explored.
%The shape of the free energy profile for WT ABL was probed with an umbrella sampling simulation,
%This was performed on an implementation of the system with a smaller simulation box and where ions where placed only to neutralize the system, 
%to provide an initial guess for the interface placements. The escape of imatinib from the bounded state is accompanied by a steep increase of the free energy well over the first 10\,\AA\ (Fig.~S2 in SI). This is indicative for imatinib being strongly bound within the binding pocket. Therefore, the distance between interfaces $\lambda_i$ needs to be small for the first 10 \AA, while it can be larger for larger $\lambda$ values. Based on this analysis, the interfaces $\lambda_i$ were chosen. The REPPTIS simulations make use of 46 interfaces covering the range $\lambda_A=1$\,\AA\ to $\lambda_B=38$\,\AA\ (see Fig.~S3 in SI) and are executed with the PyRETIS 3 software \cite{vervust2024pyretis}. The $\infty$RETIS simulations make use of 51 interfaces covering a smaller range from $\lambda_A=1$\,\AA\ to $\lambda_B=6$\,\AA\ (see Fig.~S4 in SI) and are executed with the infRETIS software \cite{infretissoftware}.

\subsection*{Trajectory-based order parameter calculation}

Calculation of $\lambda$ for a frame of the MD trajectory that is currently being generated involves (1) loading of the MD frame, (2) undoing PBC effects (making the protein and ligand whole), (3) aligning the frame onto a reference, and (4) calculating a distance. 
For such compute-heavy order parameters, the MD engine may be significantly faster than the $\lambda$-calculation. We have constructed a trajectory-based $\lambda$-calculation scheme where the order parameter is calculated for subtrajectories of MD frames. 
As most of the four subroutines have implementations that are heavily optimized for trajectories of frames rather than frame-by-frame calculations, the resulting order parameter calculation was approximately $50$-fold faster for the ABL-imatinib system. 
A detailed description of this calculation scheme is given in Text~\hyperref[sec:si]{S}2.

\subsection*{Asynchronous REPPTIS}
% infRETIS with REPPTIS implementation 'the July 2023 version'

Initially, RETIS was used to study the whole ABL-imatinib unbinding pathway using the PyRETIS 2 code \cite{riccardi2020pyretis}. However, imatinib got trapped in a metastable state in many of the trajectories, preventing imatinib to unbind in a reasonable simulation time. In practice, a maximum path 
length is set in the PyRETIS algorithm. Paths exceeding this length are forcefully rejected to restrict disk space usage, and their occurrence should be minimal.
%as they break detailed balance: actually, they don't, you limit the path ensemble to those with L<Lmax
In the RETIS simulations, nearly all
shooting moves 
or wire-fencing moves
resulted in trajectories that did not commit to $\lambda_A$ or 
$\lambda_B$ within the maximum allowed path length, even for values as high as
50\,ns. This confirmed
the presence of long-lived metastable states along the 
dissociation pathways, as predicted by the pulling and equilibrium simulations. The extremely low acceptance rate 
made RETIS infeasible and no results could be produced for the full unbinding kinetics.
The partial paths in the REPPTIS approach offer an answer to those long trajectories seen in RETIS, as the paths are cut short before they have fully committed to the bound or unbound state. 

Further large speed-up of the REPPTIS simulations was achieved by implementing 
REPPTIS in 
a custom version of 
the infRETIS software \cite{infretissoftware}. The 
$\infty$RETIS~\cite{roet2022exchanging,
zhang2024highly} algorithm
allows multiple 
\textit{workers} to run MD moves (or other MD intensive moves) in parallel, 
performing the MC moves of the path sampling algorithm asynchronously.
This implementation was intended solely for the ability to use more hardware 
resources, as the infinite swapping formalism is not applicable to 
REPPTIS ensembles. 
The software was (successfully) tested for usage with the internal PyRETIS 
engine and the external GROMACS engine. 
Our custom version allowed to run REPPTIS asynchronously on multiple GPU and CPU simultaneously, leading to a linear speed-up with computational resources.

\section*{Results}\label{sec:results}

\subsection*{REPPTIS simulations}

REPPTIS was used to study the whole imatinib unbinding pathway with the PyRETIS~3 code \cite{vervust2024pyretis}, for which the key input parameters are shown in Fig.~\hyperref[sec:si]{S}7.
%\textbf{TODO mention that we did great work about improving efficiency with the order parameter engine implementation and the asynchronous implementation.}
Order parameters were calculated every 
40 fs ($n_\text{subcycles}=20$), and the maximum path 
length was set to 4 ns. 
Occurrences of reaching the maximum path length were rare, as they only happened
[0, 0, 1, 3, 3, 25, 0, 7] times for the [WT, Q252H, Y253F, E255V, F359V,
T315I, M351T, H396P] variants. For all variants, $46$ interfaces were used, 
where the first interfaces are closely spaced (deep binding pocket) and then 
gradually spaced further apart~(Fig.~\hyperref[sec:si]{S}7). The steered 
MD simulations were used to initialize the 
REPPTIS path ensembles. 
Swap moves were attempted for $25\%$ of
the MC moves, and the remaining $75\%$ were shooting moves.

\begin{table*}[htb!]
    \centering
    \footnotesize{
    \begin{tabular}{lllllll}
    \hline
    variant                                         & $N_\text{MC}$                                   & $N_\text{ACC}$                                      & \makecell[l]{problematic\\ensembles}                            & \makecell[l]{$P_A(\lambda_B|\lambda_A)$\\$[10^{-27}]$}                    & \makecell[l]{$f_A$\\$[\text{ps}^{-1}]$}                      & \makecell[l]{$k_{\text{AB}}$\\$[\text{s}^{-1}]$} \\ \hline  
    \makecell[l]{WT\\\hphantom{a}\\\hphantom{a}}    & \makecell[l]{53967\\\hphantom{a}\\\hphantom{a}}  & \makecell[l]{15003\\\hphantom{a}\\\hphantom{a}}    & \makecell[l]{$[0-15]$, 19,\\ 25, 29, 36,\\ $[40-41]$}           & \makecell[l]{$a$\\\hphantom{a}\\\hphantom{a}}       & \makecell[l]{$4.2\cdot10^{-2}$\\$\pm19\%$\\\hphantom{a}}    & \makecell[l]{$a$\\\hphantom{a}\\\hphantom{a}}     \\
    \makecell[l]{Q252H\\\hphantom{a}}               & \makecell[l]{7156\\\hphantom{a}}                 & \makecell[l]{2499\\\hphantom{a}}                   & \makecell[l]{$[0-45]$\\\hphantom{a}}                            & \makecell[l]{$a$\\\hphantom{a}}                     & \makecell[l]{$a$\\\hphantom{a}}                       & \makecell[l]{$a$\\\hphantom{a}}     \\
    \makecell[l]{Y53F\\\hphantom{a}}                & \makecell[l]{45642\\\hphantom{a}}                & \makecell[l]{19402\\\hphantom{a}}                  & \makecell[l]{$[0-16]$, 38\\\hphantom{a}}                        & \makecell[l]{$a$\\\hphantom{a}}                     & \makecell[l]{$2.4\cdot10^{-1}$\\$\pm9.3\%$}                 & \makecell[l]{$a$\\\hphantom{a}}     \\
    \makecell[l]{E255V\\\hphantom{a}}               & \makecell[l]{72283\\\hphantom{a}}                & \makecell[l]{32597\\\hphantom{a}}                  & \makecell[l]{$[0-10]$,\\ 30, 31}                                & \makecell[l]{$1.3$\\$\pm70\%$}                            & \makecell[l]{$1.6\cdot10^{-2}$\\$\pm30\%$}                  & \makecell[l]{$2.1\cdot10^{-17}$\\$\pm76\%$}  \\
    \makecell[l]{T315I\\\hphantom{a}}               & \makecell[l]{108733\\\hphantom{a}}               & \makecell[l]{46370\\\hphantom{a}}                  & \makecell[l]{$[0-15]$,\\ $[34-36]$}                             & \makecell[l]{$32$\\$\pm53\%$}                             & \makecell[l]{$9.6\cdot10^{-3}$\\$\pm21\%$}                  & \makecell[l]{$3.1\cdot10^{-16}$\\$\pm57\%$}   \\
    \makecell[l]{M351T\\\hphantom{a}}               & \makecell[l]{44451\\\hphantom{a}}                & \makecell[l]{21481\\\hphantom{a}}                  & \makecell[l]{$[0-6]$, 17,\\ 20, 21, 27}                         & \makecell[l]{$a$\\\hphantom{a}}                           & \makecell[l]{$9.8\cdot10^{-1}$\\$\pm24\%$}                  & \makecell[l]{$a$\\\hphantom{a}}             \\
    \makecell[l]{F359V\\\hphantom{a}}               & \makecell[l]{17157\\\hphantom{a}}                & \makecell[l]{6006\\\hphantom{a}}                   & \makecell[l]{$[0-45]$\\\hphantom{a}}                            & \makecell[l]{$a$\\\hphantom{a}}                     & \makecell[l]{$1.6\cdot10^{-3}$\\$\pm85\%$}                  & \makecell[l]{$a$\\\hphantom{a}}     \\
    \makecell[l]{H396P\\\hphantom{a}\\\hphantom{a}} & \makecell[l]{62497\\\hphantom{a}\\\hphantom{a}}  & \makecell[l]{17694\\\hphantom{a}\\\hphantom{a}}    & \makecell[l]{$[0-4]$, 18,\\ 19, 32, 33,\\ $[40-42]$}            & \makecell[l]{$a$\\\hphantom{a}\\\hphantom{a}}       & \makecell[l]{$2.6\cdot10^{-3}$\\$\pm31\%$\\\hphantom{a}}    & \makecell[l]{$a$\\\hphantom{a}\\\hphantom{a}}  
    \\ \hline
    variant\vphantom{$\prod_{\prod}^{\prod}$} & \multicolumn{2}{l}{method} & \multicolumn{4}{l}{$k_{\text{off}} [\text{s}^{-1}]$} \\
    \hline
    WT~\cite{seeliger2007c}         & \multicolumn{2}{l}{experiment}    &\multicolumn{4}{l}{$(2.2\pm4.6)\cdot10^{-3}$}\\
    WT~\cite{lyczek2021mutation}    & \multicolumn{2}{l}{experiment}    &\multicolumn{4}{l}{$(8.3\pm0.8)\cdot10^{-4}$} \\
    WT~\cite{agafonov2014energetic} & \multicolumn{2}{l}{experiment}    &\multicolumn{4}{l}{$(2.5\pm0.6)\cdot10^{1}$} \\
    WT~\cite{narayan2021computer}   & \multicolumn{2}{l}{milestoning}   &\multicolumn{4}{l}{$1.8\cdot10^{1}$} \\
    WT~\cite{shekhar2022protein}    & \multicolumn{2}{l}{InMetaD
    %$^\ddagger$
    }   &\multicolumn{4}{l}{$(6\pm3)\cdot10^{-4}$} \\
    \hline
    \end{tabular}}
\caption{Results of the ABL-imatinib REPPTIS simulations (top half of table)
and rate estimates from literature (bottom half of table).
None of the simulations
have converged to a reliable rate constant. All simulations used $46$ ensembles.
$N_\text{MC}$ and $N_\text{ACC}$ denote the number of MC moves performed and 
the number of MC moves accepted, respectively. Problematic ensembles are discussed 
in the text. 
%\\\hspace{\textwidth}
$a$: Indicates that insufficient data were collected to determine the value.
%$^\ddagger$: infrequent metadynamics (InMetaD) simulation.
The provided errors were calculated using the recursive block averaging approach \cite{vervust2024pyretis} based on one standard deviation. However, for unconverged simulations, they primarily reflect variance in the data rather than providing reliable confidence intervals for the computed mean values.
}
\label{tab:ABL:reppresults}
\end{table*}

\begin{figure}[htb!]
    \includegraphics[width=0.48 \textwidth]{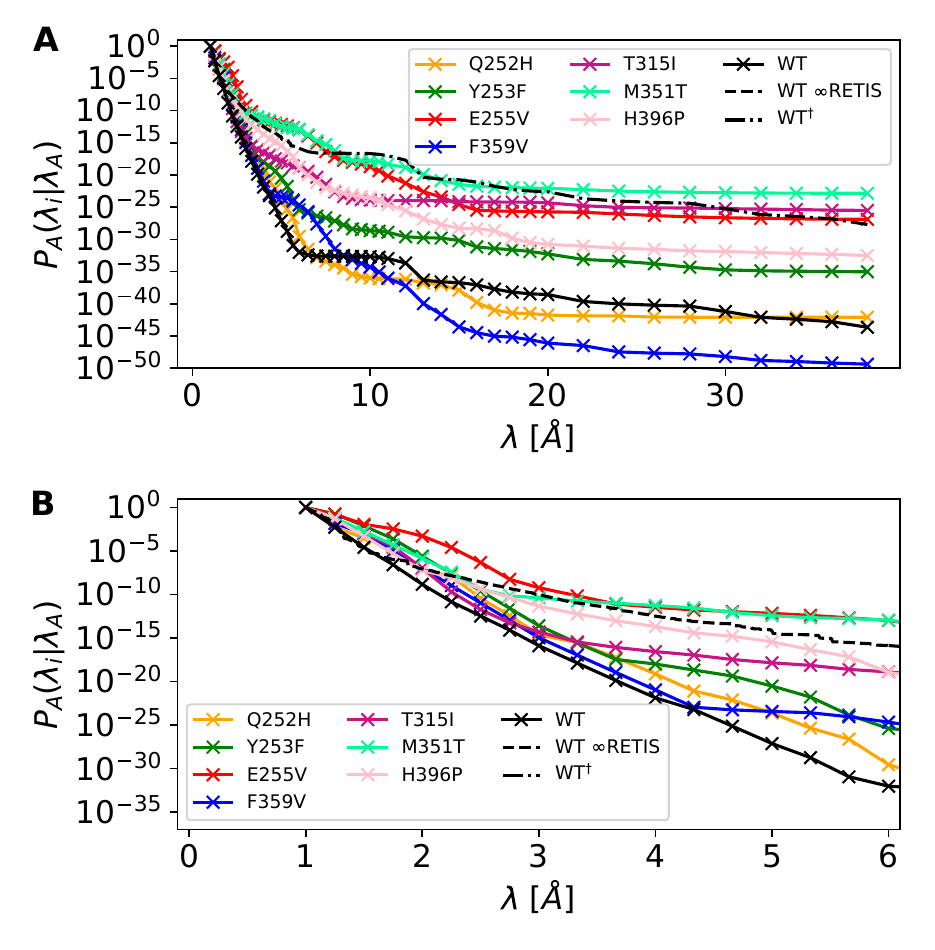}
    \caption{Crossing probability profile $P_A(\lambda_i|\lambda_A)$ for the 
    ABL-imatinib systems  (\textbf{A}) over the entire $\lambda\in[1,38]\,\text{\AA}$ 
    range and (\textbf{B}) 
    zoomed in on the $\lambda\in[1,6]\,\text{\AA}$ range.
    Profiles related to the REPPTIS simulations (lines 
    with X-markers for each interface) have artificially assigned 
    local crossing probabilities in case of zeros (see main text).
    The black dashed line denotes 
    the $\infty$RETIS simulation.
    The black dash-dotted line (WT$^\dagger$)
    is constructed by appending the $\lambda\in[6,38]\,\text{\AA}$ 
    profile of WT REPPTIS to the $\lambda\in[1,6]\,\text{\AA}$ profile of 
    WT $\infty$RETIS.
    }
    \label{fig:ABL:reppresults}
\end{figure}

The results of these simulations are shown in
Table~\ref{tab:ABL:reppresults}, where the initial path was excluded from the 
analysis.
None of the simulations converged to a reliable 
rate estimate, where most did not contain sufficient 
data
to calculate the global crossing probability $P_{A}(\lambda_B|\lambda_A)$ and 
the rate constant $k_{AB}$. 
Reliable nonzero estimates require at least one trajectory in each ensemble, centered around $\lambda_i$, that crosses $\lambda_{i+1}$ for each ensemble $i$ beyond the initialization phase. If any ensemble fails to meet this criterion, estimating the global crossing probability—and consequently the rate—becomes unfeasible, which was the case here. 
Moreover, while a rate constant 
and error estimate was obtained for the 
E255V and T315I variants, their values are not reliable
due to failed sampling in 
a large portion of the path ensembles (`problematic ensembles' column of 
Table~\ref{tab:ABL:reppresults}), which 
is discussed in more detail in the next section.
These ensembles likely failed to sample 
representative paths, resulting in extremely small (or zero-valued) local 
crossing probabilities.
Nevertheless,
to get an idea of the crossing probability profiles
$P_A(\lambda_i|\lambda_A)$,
the zero-valued local crossing probabilities were 
artificially set to $0.01$, where the complementary local crossing 
probabilities were then set to $0.99$ to keep their sum at $1$ (e.g.\ 
$p_i^\pm + p_i^{=} \equiv 1$).
The resulting profiles are shown in Fig.~\ref{fig:ABL:reppresults}.

\subsection*{Causes of REPPTIS sampling issues}

First, a note on the rate error estimations is warranted. 
While the error estimates for the E225V and T315I rates, 76\% and 57\%, give the impression of well-converged simulations, this is not the case. These rate estimates are based on the recursive block error approach~\cite{vervust2024pyretis}, and, like any other error estimators, become unreliable when the underlying data itself is poor. This is because the approach assumes convergence that may not be present, causing the estimated error to be much smaller than the actual variability. To conclude, the precision of the rate estimates is likely severely overestimated.

\begin{figure*}[htb!]
    \centering
\includegraphics[width=0.8\textwidth]{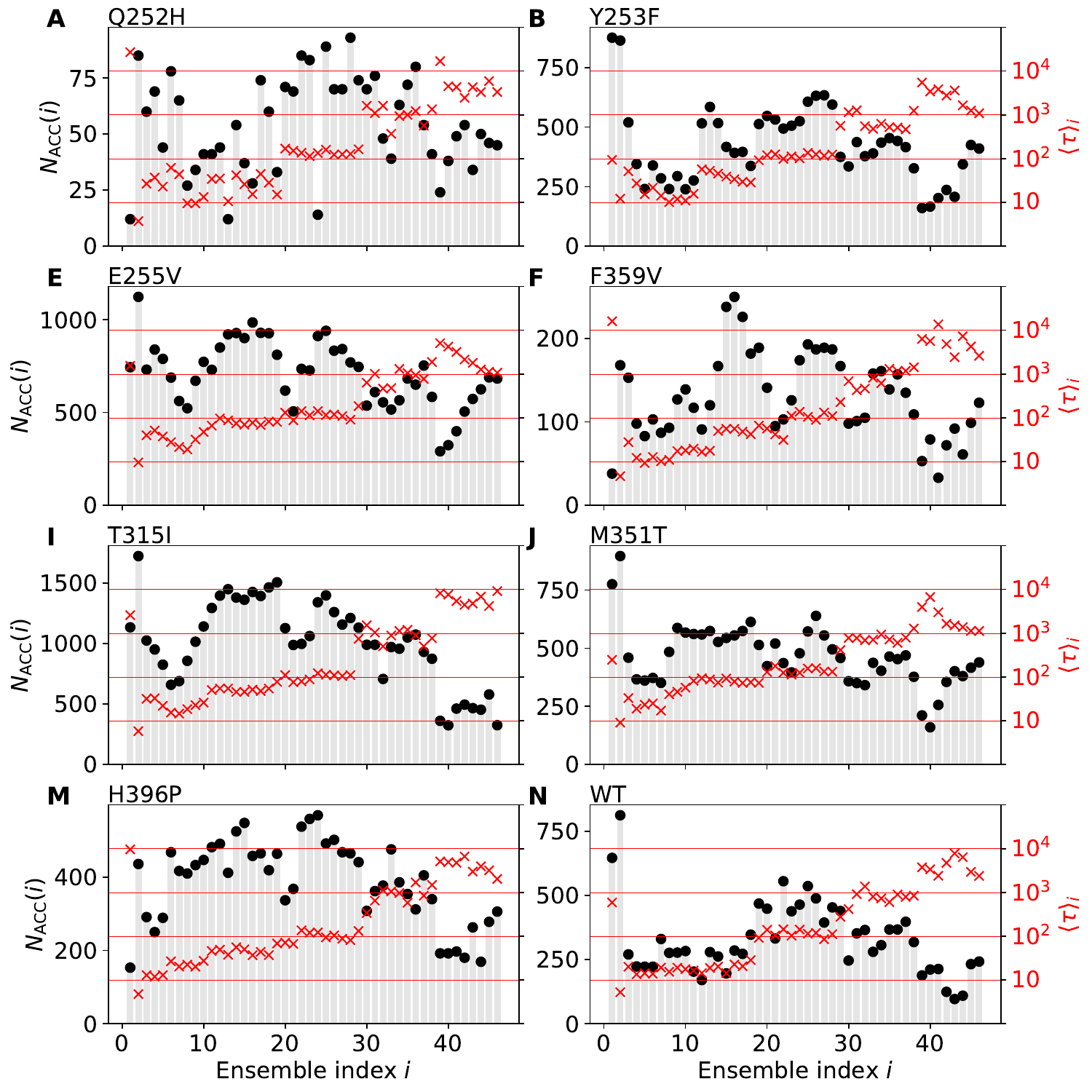}
    \caption{
    Analysis of REPPTIS simulations with 46 ensembles for each variant.
    Ensemble index $1$ corresponds to $[0^-]$,
    index $2$ to $\left[0^{\pm} \right]$, 
    index $3$ to $\left[1^{\pm}\right]$, 
    etc.
    Left y-axis: number of unique accepted paths $N_\text{ACC}(i)$ in ensemble with index $i$ is shown with 
    gray bar and black dot; different scale for different variants. 
    $N_\text{ACC}(i)$ denotes the number of accepted paths, i.e.\ the number of unique paths generated by a shooting move or replica exchange moves in REPPTIS.
    Right y-axis: average path length ${\langle \tau \rangle}_i$ is shown with red X-mark, which represents the average duration of the short unbiased trajectories in ensemble $i$. ${\langle \tau \rangle}_i$ is here measured in the number of phase points. Path phase points are separated by 40 fs, so a path length value of e.g.\ 100 represents 4\,ps. The same logarithmic scale is used for all variants.
    }
    \label{fig:ABL:Nacc}
\end{figure*}

Table \ref{tab:ABL:reppresults} includes a list of the problematic ensembles.
%The ABL-imatinib systems did not converge in the problematic ensembles (Table \ref{tab:ABL:reppresults}).
For all the simulations, these include the ensembles close to the deep 
binding pocket (small $\lambda$ values), and some ensembles further along 
the unbinding pathway. 
This was not due to a severe lack of accepted paths in these ensembles,
as there is no large discrepancy between the number of accepted paths 
$N_\text{ACC}(i)$ for the different ensembles (Fig.~\ref{fig:ABL:Nacc}). 

%Two issues are assumed to be at the root of the sampling issues. First, the Monte Carlo sampling that stimulates progress along the one-dimensional order parameter $\lambda$ fails to faciliate the exploration along orthogonal degrees of freedom. That means that, for some ensembles, energy barriers along $\lambda^\perp$ (i.e.\ the degrees of freedom orthogonal $\lambda$) can be important or even dominant for the dissociation process. Alternatively, these orthogonal barriers are in the actual transition avoided by diffusion within the reactant well along the orthogonal directions prior to progress along the $\lambda$ direction.
Two other factors are believed to contribute to the sampling challenges. First, the Monte Carlo moves designed to promote progress along the one-dimensional order parameter $\lambda$ are ineffective at facilitating exploration in orthogonal degrees of freedom. As a result, some ensembles encounter significant energy barriers along $\lambda^\perp$ (i.e., directions orthogonal to $\lambda$), which can be crucial or even dominant for the dissociation process. Alternatively, the system may bypass these orthogonal barriers during the actual transition through diffusion within the reactant well along orthogonal directions before advancing along $\lambda$.
This then 
couples with the second main issue, where the initial path may not have 
been representative of a true unbiased pathway. The orthogonal barriers then 
hinder the path sampling procedure to explore the relevant regions of 
phase space. 
As an example, it may be possible that a rotation of a dihedral angle 
between imatinib rings is the rate limiting step close to the 
deep binding pocket. If this rotation is missed in the initial path, 
then it is unlikely to be introduced by the sampling procedure if this
rotational energy barrier is large. 

These issues have been shown \cite{van2006efficiency} to be substantially less pronounced in transition interface sampling (TIS) compared to other methods relying on a one-dimensional order parameter, such as forward flux sampling (FFS) \cite{allen2009forward}, umbrella sampling (US), metadynamics with a single collective variable, or methods based on a dividing surface, like the Bennett-Chandler method \cite{frenkel2023understanding}. The incorporation of replica exchange in RETIS further mitigates these challenges. For instance, RETIS successfully captured distinct parallel transition channels in a permeation study \cite{Ghysels2021exactnonMarkov}, where orthogonal diffusion was enhanced using tailored Monte Carlo moves in the $[0^-]$ ensemble, specifically designed for such systems. These moves significantly improved sampling efficiency across all ensembles by leveraging the interconnectedness of replica exchange moves.
However, the shorter trajectories in PPTIS are inherently more local and thus more prone to getting trapped than those in TIS and RETIS. Additionally, the replica exchange moves in REPPTIS are evidently insufficient to fully resolve this issue, highlighting the need for further enhancements to address orthogonal barriers.

The average path length $\langle \tau \rangle_i$ of each ensemble $i$, $\forall i\in\{1,..,46\}$, 
is also shown in 
Fig.~\ref{fig:ABL:Nacc}, for each of the variants. The path length
is measured in `number of phasepoints', where the starting and ending 
phase points of paths are excluded, and the average takes into account the weight of the path in the MC procedure.
The path lengths are much lower near the binding pocket, which is to be 
expected as the energy barrier is steepest in this region, and trajectories easily fall back into the reactant state.
%As the occurrence of LMR or RMR paths is rare in these ensembles,
As the occurrence of paths that reach their right-most interface is rare in these ensembles, 
the REPPTIS swap move, which is a replica exchange MC move designed to improve phase space exploration, was
rarely performed.

Let us now focus on WT ABL, whose crossing probability profile
$P_A(\lambda_i|\lambda_A)$ showed the fastest drop of all the variants 
(Fig.~\ref{fig:ABL:reppresults}).
Path lengths for the WT ABL system in the 
$[1^+], \dots, [16^+]$ ensembles 
(ensemble indices $3$ to $18$ on Fig.~\ref{fig:ABL:Nacc}N) vary around $20$, 
with approximately $250$ accepted paths per ensemble. This
corresponds to 
$20\times 40$  fs $\times 250 = 200$ ps of combined 
trajectory lengths, for each of these ensembles. Only a fraction of this time 
can be seen as `equilibration time', as the shooting points are (on average) not 
located near the beginning or end of paths. It is suspected that the initial path 
of WT ABL was not representative for the dominant reaction pathway, 
where the steered MD simulation 
pulled it along a barrier orthogonal to $\lambda$.
This barrier was then not overcome by the fraction of the 200 ps 
of phase space exploration. Therefore, the local 
crossing probabilities $p^\pm_i$ in these ensembles remain 
extremely low (or zero).
For problematic ensembles further along the unbinding pathway 
($\left[i^{\pm}\right]$, with $i\geq 20$), the path lengths are considerably longer.

Moreover, the efficiency of phase space and path space exploration is influenced not only by the path lengths but also by correlations between successive trajectories. These correlations arise because successive trajectories are likely confined to the same reaction channel, a consequence of the high orthogonal energy barriers. For these ensembles, such orthogonal barriers are expected to hinder efficient sampling of the corresponding $\left[i^{\pm}\right]$ path spaces. 
For the 6 mutants with initial paths dissociating under the $\alpha$C-helix (Text~\hyperref[sec:si]{S}3, Fig.~\hyperref[sec:si]{S}3), this means that a converged simulation would likely only capture the rate associated with escape under the $\alpha$C-helix, and not a `net' rate taking into account possible dissociation pathways under/along the P-loop. 
This reasoning extends to the other mutated variants as well but is expected to be less pronounced in TIS and even less so in RETIS and $\infty$RETIS, where enhanced sampling techniques mitigate these challenges.

\subsection*{$\infty$RETIS simulations}\label{sec:ABL:infretis}

To further investigate the sampling issues close to the binding pocket, 
an $\infty$RETIS simulation was run for the WT ABL system for the 
$\lambda\in[1,6]\,\text{\AA}$ range, focusing on the steep free energy increase associated to the first part of the imatinib unbinding pathway.
The input parameters used for the $\infty$RETIS simulation of WT ABL
are shown in Fig.~\hyperref[sec:si]{S}8.
Order parameters were
calculated every 1~ps ($n_\text{subcycles}=500$), and the maximum path length was set
to 100 ns.
Interfaces were placed from $\lambda_A =1$ \AA\ to
$\lambda_B =6$ \AA\ with a separation of 0.1 \AA,
resulting in $51$ interfaces and $51$ ensembles 
$[0^-], [0^+], [1^+], \dots,  [49^+]$. 
The interfaces can be put in closer proximity 
to each other, as the infinite swapping formalism of $\infty$RETIS efficiently 
distributes (swaps) path information between the ensembles.
Furthermore, the optimal number of workers 
(MD intensive moves that can run in parallel) is approximately half of the 
number of ensembles,
where it is thus 
advantageous to place more interfaces than
what is standard for a PyRETIS simulation using the RETIS algorithm.

\begin{figure}[htb!]
    \includegraphics[width=0.48\textwidth]{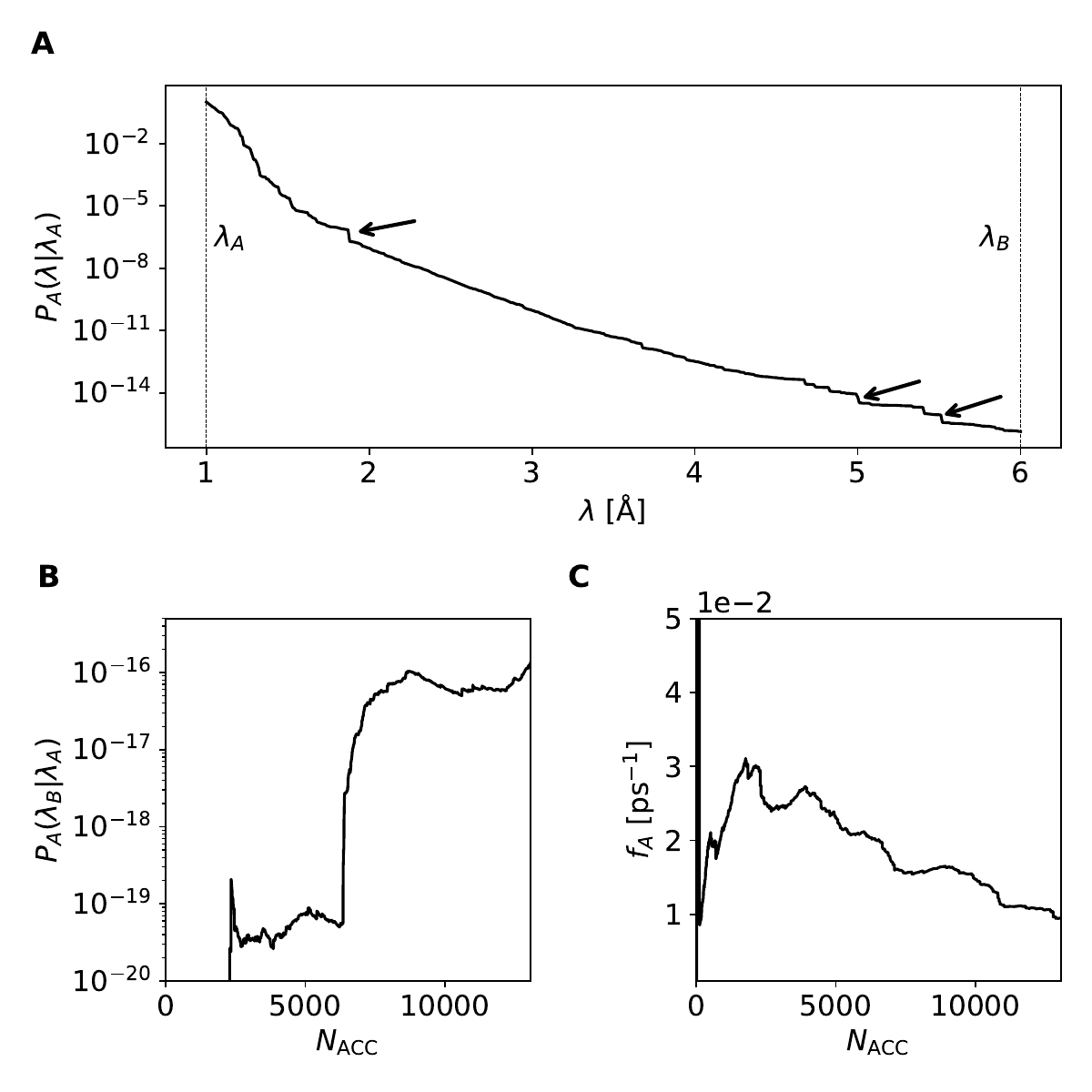}
    \caption{WT ABL $\infty$RETIS results, where the first 5000 accepted paths 
    were discarded to reduce initialization effects. 
    \textbf{A}: The crossing probability profile $P_A(\lambda|\lambda_A)$ of Eq.~\ref{eq:rate}. 
    Arrows indicate visible discontinuities.
    \textbf{B}: The running estimate of the global crossing probability 
    $P_A(\lambda_B|\lambda_A)$ shows a large jump, hinting at non-convergence of the simulation.
    \textbf{C}: The running estimate 
    of the flux $f_A$ of Eq.~\ref{eq:rate}. $N_\text{ACC}$ denotes the number of accepted paths i.e.\ the number of unique paths generated by a shooting move. 
    }
    \label{fig:ABL:infretiswham}
\end{figure}

The resulting crossing probability profile $P_A(\lambda|\lambda_A)$ 
is included in Fig.~\ref{fig:ABL:reppresults}A-B (black dashed line) 
for comparison with the REPPTIS profiles, and also shown separately in
Fig.~\ref{fig:ABL:infretiswham}A.
The first $5000$ of the $18\,070$ accepted paths were discarded in the analysis to reduce 
initialization effects. 
The weighted histogram analysis (WHAM) method approach of RETIS produces a continuous crossing probability profile, rather than 
discrete points at the interface positions for REPPTIS. 
The drop in crossing probability is 
dramatically 
less steep than the WT REPPTIS profile with about 16 orders of magnitude.
However, discontinuities in the 
$P_A{(\lambda|\lambda_A)}$ profile are visible (arrow indications on 
Fig.~\ref{fig:ABL:infretiswham}A), and are discussed next.
%@lambda=6:
%repptis: 9.65e-33
%retis: 1.38e-16

%is still expected to be exaggerated, 
%especially in the $\lambda\in[1,2]\,\text{\AA}$ range. 
%{\bf compared to?? ...}, 
%Compared to 1) hindsight fact that the probability must be a lot lower (and these ensembles are sampled the worst)
%2) The crossing probability curve in lambda \in [1,2]angstrom is still very similar to the REPPTIS ones, which are also underestimated  (just like for trypsin benzamidine, which we could verify with plain MD) 

\begin{figure}[htb!]
    \centering
    \includegraphics[width=0.480\textwidth]{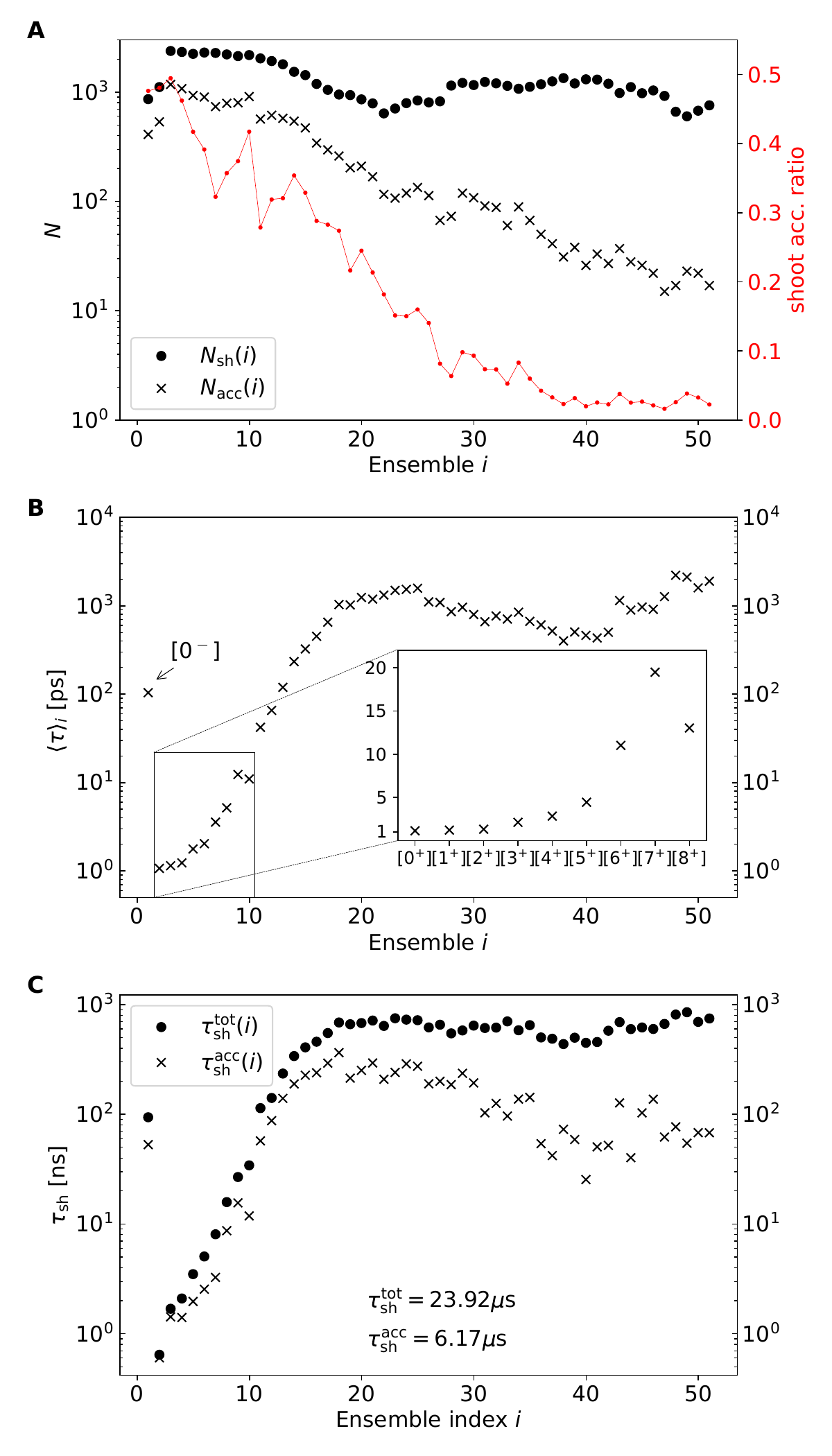}
    \caption{Sampling analysis of the $\infty$RETIS WT ABL simulation.
    All \textit{shooting} cycles are included in the analysis,
    i.e.\ paths generated by the
    zero swap move in 
    $[0^-]$  and $[0^+]$
    are excluded. 
    Ensemble index 1 corresponds to $[0^-]$, ensemble index 2 to $[0^+]$, ensemble index 3 to $[1^+]$, etc.
    \textbf{A}: $N_\text{sh}(i)$ is the number of performed shooting moves in the $\infty$RETIS simulation, both accepted and rejected (black dots). $N_\text{acc}(i)$ is the number of accepted shooting moves (black crosses). The acceptance ratio is shown in red.
    \textbf{B}: Average path length ${\langle\tau\rangle}_i$, representing the ensemble average duration of the short unbiased paths of ensemble $i$. 
    %Extremal phase points of paths are excluded.
    \textbf{C}: The total amount of simulated (shooting) time $\tau_\text{sh}(i)$
    (dots)
    and the accepted portion 
    $\tau^\text{acc}_\text{sh}(i)$ thereof (crosses). Path length is measured as the number of phase points of the path, which are saved every 1\,ps. E.g.\ a path length of 100 corresponds to 100\,ps.
    }
    \label{fig:ABL:infretisplots}
\end{figure}

The discontinuity around $\lambda\approx 1.9$ {\AA} is due to
the undersampling of far-reaching trajectories in the first $9$ positive ensembles
$[0^+], \dots, [8^+]$. The average paths lengths of these ensembles 
(now measured in ps, as 
$n_\text{subcycles}\times\text{timestep}=
1$ ps)
is too small to promote equilibration 
by phase space exploration (Fig.~\ref{fig:ABL:infretisplots}B). 
Most of the paths in these ensembles consist 
of only a single phase point. 
Algorithmically, this is equivalent to paths containing
three phase points, 
where the first and last points are not a part of the path ensemble (they 
belong
to state A or B) and therefore excluded from being viable 
shooting points. 
This means that, for these $1$-phase point paths, 
the configuration remains identical
as only the momenta are modified. 
It is clear that decorrelation between paths is extremely slow when there is only one phase point available to shoot from.
%Thus, although many paths were accepted in these ensembles (Fig.~\ref{fig:ABL:infretisplots}A), the exploration of the ensemble's path space remains extremely slow due to the high degree of correlation among the shooting points.
Thus, despite the high acceptance rate of paths in these ensembles (Fig.~\ref{fig:ABL:infretisplots}A), exploration of the ensemble's path space remains sluggish due to the strong correlation among shooting points.

The discontinuities in \( P_A(\lambda|\lambda_A) \) for \(\lambda \geq 4.5 \, \text{\AA}\) arise due to the limited number of accepted paths (\(\leq 100\)) in the corresponding ensembles (Fig.~\ref{fig:ABL:infretisplots}A), despite each ensemble having simulated approximately 500 ns or more of shooting time (Fig.~\ref{fig:ABL:infretisplots}C). This low acceptance could have been mitigated using the wire-fencing move with high-acceptance~\cite{zhang2023enhanced}, a recent advanced shooting move designed to improve decorrelation and achieve near rejection-free sampling, particularly for ensembles close to the reactant state. Such a move would likely enhance the exploration of under-sampled path spaces and improve the convergence of the corresponding ensembles. However, we will investigate in future studies whether this approach also resolves issues arising from occurrences when there is only one phase point to shoot from.

Based on the observed variation in path lengths, it is clear that metastable states are present in the WT ABL-imatinib system, even for order parameter values as low as $\lambda = 2 \, \text{\AA}$. This suggests that the separation of timescales introduced by these states might fundamentally \textit{not} be resolvable by a one-dimensional $\lambda$-based approach. Specifically, a return to state $A$ from $\lambda \geq 2 \, \text{\AA}$ can occur within just a few RETIS timesteps of 1 ps—or even a single timestep—when no metastable state is encountered. Conversely, when the trajectory becomes trapped in a metastable state, returning to state $A$ can take thousands of timesteps. This highlights the inherent limitations of relying on a single reaction coordinate to capture the complexities of the dynamics involved. 
The importance of the order parameter choice on simulation convergence is further discussed in the \hyperref[sec:discussion]{Discussion} section, together with alternative order parameter definitions that may provide better performance for complex protein-drug systems like ABL-imatinib.

Although $\infty$RETIS shows a significant improvement over REPPTIS, and even more so over PPTIS, in handling a single $\lambda$ parameter for enforcing progress, and despite the potential for further efficiency gains with advanced shooting moves, the very large average path lengths observed for ensembles beyond $6$\,\AA\ make this approach currently unfeasible. A (RE)PPTIS or milestoning approach seems more practical under these conditions.

A strategy could be to combine both methods, in order to leverage the better sampling efficiency of $\infty$RETIS at the start and the shorter path length of REPPTIS further along the reaction. As indicated by the black dash-dotted line in Fig.~\ref{fig:ABL:reppresults}), we extended the 
$\infty$RETIS crossing 
probability profile with the $\lambda \in [6, 38]\text{\AA}$ range from the WT REPPTIS simulation. This gives a global crossing probability \( P_A(\lambda_B|\lambda_A) \) of $2.08\times10^{-28}$. When combined with the WT ABL flux from $\infty$RETIS ($9.51\times10^{-3}$ ps$^{-1}$), the resulting rate estimate for imatinib dissociation from WT ABL is $1.98\times10^{-18}$ s$^{-1}$. 
%This estimate, despite being 16 orders of magnitude higher than the pure REPPTIS results, remains more than 10 orders of magnitude too low 
%different from other 
%experimental 
%or computational rate 
%measurements.
This estimate, despite being 16 orders of magnitude higher than the pure REPPTIS result, remains more than 10 orders of magnitude lower than experimental measurements.

It is noteworthy that the variation among experimental results, as summarized in Table \ref{tab:ABL:reppresults}, spans nearly five orders of magnitude,
and also
simulation-based approaches, based on milestoning and infrequent metadynamics, differ by nearly five orders of magnitude. Although these variations are large, they are still significantly closer to each other than to the combined $\infty$RETIS-REPPTIS estimate. 
The large underestimation of the crossing probability in our simulation is likely not only the result of force field discrepancies, but more importantly the result of sampling issues, as discussed below.
 
%----------------------------------
\section*{Discussion}\label{sec:discussion}
%----------------------------------

Both RETIS and REPPTIS faced significant challenges in achieving well-converged kinetics for imatinib dissociation from ABL. These challenges most likely stem from the complexity of the unbinding mechanism, where the Monte Carlo exploration of path space struggles to adequately sample along directions orthogonal to the reaction coordinate $\lambda$. This orthogonal sampling is crucial because spontaneous dynamical transitions may involve crossing orthogonal barriers, leading to alternative pathways with potentially lower barriers along $\lambda$. Although these routes might not initially appear as paths of least resistance, they can ultimately result in more favorable dissociation trajectories.

For REPPTIS, barriers orthogonal to $\lambda$ prevent paths from equilibrating within the relevant regions of path space. If the initial path segments (paths truncated to the $[i^\pm]$ ensemble boundaries) are not representative of a true dissociating pathway, the short path lengths and the strict locality of the path ensemble definitions hinder the exploration of relevant path space, particularly for ensembles near the binding pocket. This issue is exacerbated by the low acceptance of replica exchange moves, which are crucial for facilitating transitions along orthogonal degrees of freedom. The steepness of the orthogonal energy barriers results in most trajectories returning to $\lambda_{i-1}$ after reaching $\lambda_{i}$ from $\lambda_{i-1}$, with few successfully advancing to $\lambda_{i+1}$. 
Since paths returning to $\lambda_{i-1}$ cannot be swapped with a path in the $[(i+1)^\pm]$ ensemble, the replica exchange acceptance rate is low.

%While RETIS, and more so $\infty$RETIS, 
While $\infty$RETIS
improved sampling in the $\lambda \in [2,6]$\,{\AA} region, the separation of timescales necessitates a significantly smaller timestep (as defined by $n_\text{subcycles}$, the number of MD steps that comprise a full TIS timestep) for effective sampling in the $\lambda \in [1,2]$\,{\AA} region. This higher saving frequency would result in substantially larger disk space usage and extended simulation times due to increased data-writing overhead.
%This would demand substantial disk space, as trajectories at higher $\lambda$ values often exceeded 60\,ns. The high frame-saving rate required to capture such dynamics would further extend simulation times due to increased data-writing overhead.
An even larger problem is that setting $\lambda_B = 6$\,{\AA} does not ensure full dissociation. Increasing $\lambda_B$ would introduce exceptionally long paths as imatinib is frequently (transiently) trapped in metastable states between the binding pocket and solvent.  %Methods like (RE)PPTIS or milestoning are better equipped to handle these dynamics which are both rare \emph{and} slow.
Methods such as (RE)PPTIS and milestoning are better suited for handling dynamics that are both rare and slow.
%However, shorter paths are more prone to becoming trapped, 
However, the validity of the Markovian assumption implies by (RE)PPTIS depends on the choice of order parameter.

%New paragraph on alternatvie order parameter definitions.
%The selection of an appropriate order %parameter $\lambda$ is crucial
%in path sampling simulations, as it 5significantly influences the %convergence rate and computational %efficiency for the RE(PP)TIS %simulations. This holds true also for %trajectory splitting methods such as 
%milestoning %\cite{faradjian2004computing,ojha2024adv%ances}, weighted ensemble method %\cite{huber1996weighted}, and adaptive %multilevel splitting %\cite{cerou2011multiple}. 
%****
%reformulation:
The selection of an appropriate order parameter $\lambda$ is critical for all enhanced sampling methods, as it can influence both computational efficiency and the results upon convergence. For (RE)PPTIS and Milestoning, an unfavorable choice of $\lambda$ can compromise 
%the validity of the Markovian assumption,
Markovianity,
where results diverge from those expected from (extremely long) MD simulations. 
%In contrast, other methods will, in principle, yield exact results regardless of $\lambda$; however, a poor choice of $\lambda$ can result in an exponential increase in computational cost. 
While other methods will, in principle, yield exact results regardless of $\lambda$, a poor choice of $\lambda$ can result in an exponential increase in computational cost.
TIS and RETIS, in this respect, theoretically offer an advantage over splitting-based methods such as 
forward flux sampling, weighted ensemble method \cite{huber1996weighted}, and adaptive multilevel splitting \cite{cerou2011multiple} and Bennett-Chandler-type approaches, as their efficiency is less sensitive to the order parameter. This advantage has been demonstrated for simple 1D and 2D model systems \cite{van2012dynamical,van2006efficiency}. However, in complex, high-dimensional systems, the choice of $\lambda$ is likely to remain crucial, even for these methods.

When slow (diffusive) or rare (energetic) barriers important to the reaction mechanism are not captured by $\lambda$, then convergence can become drastically problematic, as was the case for the simple one-dimensional distance $\lambda$ parameter used in this ABL-imatinib study.
Alternative one-dimensional $\lambda$ parameters may be better suited for complex protein-drug dissociation studies. 
A contact map parameter based on hydrogen bond donor/acceptor distances between the protein and drug may capture important steps in the dissociation mechanism that are not accompanied by a simple displacement from the native binding pose.   
Such a parameter was successfully used in studying the association and dissociation of protein-DNA complexes~\cite{riccardi2019predicting,van2023decoding} and may also be better suited here. 
%The possibility of multiple dissociation pathways with multiple metastable states sets a challenging task for the creation of a one-dimensional order parameter.
Another alternative is to employ path collective variables (PCVs) that tackle a specific dissociative pathway for each of the  ABL mutants. Such a pathway can be constructed via the string method~\cite{weinan2002string,maragliano2006string}, or via machine-learning methodologies that, for example, employ autoencoders to construct a PCV using a low-dimensional representation of the system
\cite{frassek2021extended,belkacemi2021chasing,frohlking2024deep}.
The ABL-imatinib simulations in this work showed that the addition of replica exchange to the PPTIS framework was unable to alleviate the crucial dependency of the convergence speed on $\lambda$, and going forward it will be rewarding to allocate compute time on suitable $\lambda$ design prior to running the costly RE(PP)TIS drug dissociation simulations.  

Long MD simulations on the special purpose Anton~2 supercomputer 
have revealed that there are a multitude of 
long-lived metastable states, even close to and within the 
deep binding pocket~\cite{paul2020diversity, ayaz2023structural}. 
%(Roux et.\ al.~\cite{paul2020diversity}, 
%Shaw et.\ al.~\cite{ayaz2023structural}). 
While the presence of metastable states is not problematic for REPPTIS, kinetic 
analysis of the ABL-imatinib system in the recent works of 
Refs.~\cite{narayan2021computer,shekhar2022protein}
suggested that the dissociation process 
 requires
a multi-dimensional 
description in collective variable spaces and that a one-dimensional order parameter is fundamentally not sufficient.
%might not be fundamentally well described by a one-dimensional reaction coordinate.
A multidimensional
milestoning method was used by 
Elber et al.~\cite{narayan2021computer}, where Voronoi tessellation revealed 
a milestoning network with average connectivity of $2.93$, whereas a value of 
$2$ is expected for a one-dimensional description. Infrequent metadynamics 
simulations by Shekhar et al.~\cite{shekhar2022protein} used a 5-dimensional 
reaction coordinate model to extract the rate constant of the ABL-imatinib system. The results 
of these studies were also included in Table~\ref{tab:ABL:reppresults}, and 
are in much closer agreement with the few experimental values available. 
The study of Shekhar et al.~\cite{shekhar2022protein} also found that, in wild type ABL, imatinib predominantly dissociates through a route near the kinase hinge region, which is a pathway that becomes accessible following the disruption of a hydrogen bond between Y253 (from the P-loop) and N322 (from the hinge). In our study, the initial steered MD pathway generated for wild type ABL escaped under the $\alpha$C-helix (Fig.~\hyperref[sec:si]{S}3), where the aforementioned hydrogen bond is not disrupted. As this hydrogen bond does not directly impact the order parameter (i.e.\ it is orthogonal to $\lambda$), its disruption is hard to sample in REPPTIS.
Similarly, changes in characteristic motifs of (mutated) ABL structures are expected not to be captured by $\lambda$, resulting in impeded path sampling. 

To summarize,
while the Monte Carlo approach to path generation is a key strength of REPPTIS as initial paths naturally evolve to representative paths via importance sampling, this `path equilibration' process is impeded when orthogonal barriers are too high, and more so when replica exchange is frustrated due to a steep slope along $\lambda$. 
While improved initial path generation might have enhanced the convergence of the ABL-imatinib REPPTIS simulations, it may also indicate that REPPTIS needs to be generalized to a multidimensional order parameter to address this and similarly complex biological systems.

\section*{Future perspectives and conclusion}\label{sec:conclusion}

Future work will focus on combining REPPTIS with multiple-state TIS~\cite{rogal2008multiple}, where metastable state definitions could be derived from a Voronoi tessellation approach similar to that used in directional milestoning or from recent machine learning algorithms like Vampnets~\cite{wu2020variational}. In particular, partially retaining the memory effect inherent to REPPTIS may be advantageous by relaxing the requirements for defining anchoring points in the tessellation or the boundaries of metastable states in collective variable space. This is because, in the PPTIS formalism, transitions between states do not need to be strictly Markovian.

In addition, it is interesting to increase the memory so that the terminating boundaries in the ensemble centered around $\lambda_{i}$ are not necessarily limited to the neighboring interfaces $\lambda_{i+1}$ and $\lambda_{i-1}$, but can extend to interfaces that are further away. This concept has already been tested in simple toy systems \cite{vervust2024Thesis} but could also be combined with a multidimensional order parameter network.
Such an approach could eventually provide a seamless connection between $\infty$RETIS and REPPTIS, beyond the \textit{ad hoc} method used here. This would allow path ensembles to adapt and treat different characteristics of the free energy landscape in a way that aligns more closely with either $\infty$RETIS or REPPTIS, depending on which method is best suited for the underlying free energy landscape.

We also plan to enhance the shooting move, where the high acceptance rates and rapid decorrelation of wire-fencing is preferable for future studies. 
%Another avenue we plan to explore is enhancing the shooting move. Wire-fencing has demonstrated high acceptance rates and rapid decorrelation, making it preferable over shooting in future studies. 
However, our results suggest an alternative approach for high-dimensional systems. 
High frame-saving rates slow computation and increase disk space demands, while low rates limit the number of shooting points. 
%Using a high frame-saving rate can lead to issues with writing large amounts of data (several gigabytes), which is slow and computationally expensive. Additionally, calculating the order parameter $\lambda$ can be costly. On the other hand, a low frame-saving rate limits the number of shooting points. 
Ideally, the frame-saving rate should be flexible. Wire-fencing could provide an elegant solution by incorporating a high frame-saving rate for the sub-trajectories, which are not sampled but serve as intermediates between the actual paths in the Monte Carlo approach, while maintaining a lower frame-saving rate for the actual paths. 

The orthogonal sampling could be further enhanced by running metadynamics simulations in parallel and incorporating replica exchange moves between metadynamics configurations and randomly selected shooting points from trajectories~\cite{Falkner2024Enhanced}. This approach has the advantage that the $\lambda$ parameter can remain low-dimensional, and in some cases one-dimensional, since the multidimensional diffusion and orthogonal barrier crossings primarily occur during the metadynamics simulations. The effects of these crossings are then transferred to the path ensembles via the swapping moves.

In summary, while the challenges posed by orthogonal barriers and high-dimensional systems remain significant, the methodologies discussed here offer promising pathways to enhance sampling and facilitate the exploration of complex molecular processes. Future advancements, such as the integration of multidimensional order parameters and improved sampling techniques like wire-fencing and parallel metadynamics, are poised to refine these approaches further. By leveraging these strategies, we can advance our ability to capture rare events and transition pathways, ultimately contributing to a deeper understanding of molecular dynamics and complex biological systems.

%combining metadynamics and path sampling in a replica 

%\section{Thursday 29 August, to be included in conclusion or add a discussion}

%How to avoid being stuck in irrelevant region. Not enough swaps.
%(That is likely the only problem for benzene)

%Aspects still using the 1D coordinate

%1a) (1D situation) bad sampling along orthog coord, not even because of a barrier.
%Solution maybe: change coordinate frequency saving rate in wire fencing

%1b) investigate RC before getting started. Closer to committor.

%1c) what if started in a non-relevant channel. Initialization should be good.

%1d) bad sampling RETIS high ensembles, too many B-to-B paths. With wf, it would work better, if you have a lambda-cap. You would stop the fire fencing pieces befoe spending too much time falling all the way down into B when you shoot from a phase point that is too close to B.

%Alternative: What if going to higher dimensions?

%2) metastable states all over the place.

%infrequent metadynamics vs milestoning: ratio rates is 30 000

%Solution: do importance sampling other coordinates as well. 

%Higher-dimensional space, 1 lambda not enough, Voronoi cells, kind of multiple states but not necessarily, detect other important lambda's with VAMPnets

\phantomsection{}\label{sec:si}
\subsection*{Supplemental Information}

Additional details are provided in the Supplemental Information, including Figures S1-S6, Table S1, and Texts S1-S3.
% VMD of PDBs
% list of rigid carbon alpha atoms, table
% umbrella sampling simulations,
% key parameters of the simulations REPPTIS
% key parameters of the simulations infRETIS

\subsection*{Code availability}

%TODO Wouter December 6, 2024.

Analysis was performed using \href{www.pyretis.org}{PyRETIS} and \href{https://github.com/infretis/inftools}{inftools}. 
The custom version of $\infty$REPPTIS is available on request. The configurations and topology files of the ABL-imatinib systems are available on Zenodo~\cite{zeenoodoo_abl}.
%, and custom scripts available on github/WouterWV/...

\subsection*{Author contributions}

WV developed the simulation protocol, ran simulations, and performed the analysis. 
DTZ and ER assisted in software development and running of simulations.
TvE assisted in developing the simulation protocol.
AG was responsible for the conceptualization and supervision.
WV, TvE, and AG wrote the paper.
All authors reviewed the paper.

\subsection*{Declaration of interests}

The authors declare no competing interests.

\subsection*{Acknowledgments}

The computational resources (Stevin Supercomputer Infrastructure) and services used in this work were provided by the VSC (Flemish Supercomputer Center), funded by Ghent University, FWO and the Flemish Government – department EWI. AG acknowledges funding of the FWO (project G002520N and project G094023N) and the European Union (ERC Consolidator grant, 101086145 PASTIME).

%================================================================
%\section*{Appendix}
%================================================================

% bibstyle 
%\bibliographystyle{plain}
%\bibliographystyle{apsrev4-2}
\bibliography{BPJreferences}

\begin{thebibliography}{97}
\providecommand{\url}[1]{\texttt{#1}}
\providecommand{\urlprefix}{ }

\bibitem[Lipinski et~al.(2012)Lipinski, Lombardo, Dominy, and Feeney]{lipinski2012experimental}
Lipinski, C.~A., F.~Lombardo, B.~W. Dominy, and P.~J. Feeney, 2012.
\newblock Experimental and computational approaches to estimate solubility and permeability in drug discovery and development settings.
\newblock \emph{Adv. Drug Deliv. Rev.} 64:4--17.

\bibitem[Jorgensen(2004)]{jorgensen2004many}
Jorgensen, W.~L., 2004.
\newblock The many roles of computation in drug discovery.
\newblock \emph{Science} 303:1813--1818.

\bibitem[Claveria-Gimeno et~al.(2017)Claveria-Gimeno, Vega, Abian, and Velazquez-Campoy]{claveria2017look}
Claveria-Gimeno, R., S.~Vega, O.~Abian, and A.~Velazquez-Campoy, 2017.
\newblock A look at ligand binding thermodynamics in drug discovery.
\newblock \emph{Expert Opin. Drug Discov.} 12:363--377.

\bibitem[Mobley and Gilson(2017)]{mobley2017predicting}
Mobley, D.~L., and M.~K. Gilson, 2017.
\newblock Predicting binding free energies: frontiers and benchmarks.
\newblock \emph{Annu. Rev. Biophys.} 46:531--558.

\bibitem[Copeland et~al.(2006)Copeland, Pompliano, and Meek]{copeland2006drug}
Copeland, R.~A., D.~L. Pompliano, and T.~D. Meek, 2006.
\newblock Drug--target residence time and its implications for lead optimization.
\newblock \emph{Nat. Rev. Drug Discov.} 5:730--739.

\bibitem[Tummino and Copeland(2008)]{tummino2008residence}
Tummino, P.~J., and R.~A. Copeland, 2008.
\newblock Residence time of receptor- ligand complexes and its effect on biological function.
\newblock \emph{Biochem.} 47:5481--5492.

\bibitem[Copeland(2010)]{copeland2010dynamics}
Copeland, R.~A., 2010.
\newblock The dynamics of drug-target interactions: drug-target residence time and its impact on efficacy and safety.
\newblock \emph{Expert Opin. Drug Discov.} 5:305--310.

\bibitem[Lu and Tonge(2010)]{lu2010drug}
Lu, H., and P.~J. Tonge, 2010.
\newblock Drug--target residence time: critical information for lead optimization.
\newblock \emph{Curr. Opin. Chem. Biol.} 14:467--474.

\bibitem[Copeland(2016)]{copeland2016drug}
Copeland, R.~A., 2016.
\newblock The drug--target residence time model: a 10-year retrospective.
\newblock \emph{Nat. Rev. Drug Discov.} 15:87--95.

\bibitem[Wang et~al.(2011)Wang, McLeod, and Weinshilboum]{wang2011genomics}
Wang, L., H.~L. McLeod, and R.~M. Weinshilboum, 2011.
\newblock Genomics and drug response.
\newblock \emph{New England Journal of Medicine} 364:1144--1153.

\bibitem[Relling and Evans(2015)]{relling2015pharmacogenomics}
Relling, M.~V., and W.~E. Evans, 2015.
\newblock Pharmacogenomics in the clinic.
\newblock \emph{Nature} 526:343--350.

\bibitem[Sneha and {George Priya Doss}(2016)]{sneha2016molecular}
Sneha, P., and C.~{George Priya Doss}, 2016.
\newblock Chapter Seven - Molecular Dynamics: New Frontier in Personalized Medicine.
\newblock \emph{In} R.~Donev, editor, Personalized Medicine, Academic Press, volume 102 of \emph{Advances in Protein Chemistry and Structural Biology}, 181--224.

\bibitem[Georgoulia et~al.(2019)Georgoulia, Todde, Bjelic, and Friedman]{georgoulia2019catalytic}
Georgoulia, P.~S., G.~Todde, S.~Bjelic, and R.~Friedman, 2019.
\newblock The catalytic activity of Abl1 single and compound mutations: Implications for the mechanism of drug resistance mutations in chronic myeloid leukaemia.
\newblock \emph{Biochimica et Biophysica Acta (BBA)-General Subjects} 1863:732--741.

\bibitem[O'hare et~al.(2012)O'hare, Zabriskie, Eiring, and Deininger]{o2012pushing}
O'hare, T., M.~S. Zabriskie, A.~M. Eiring, and M.~W. Deininger, 2012.
\newblock Pushing the limits of targeted therapy in chronic myeloid leukaemia.
\newblock \emph{Nat. Rev. Cancer} 12:513--526.

\bibitem[Stegmeier et~al.(2010)Stegmeier, Warmuth, Sellers, and Dorsch]{stegmeier2010targeted}
Stegmeier, F., M.~Warmuth, W.~Sellers, and M.~Dorsch, 2010.
\newblock Targeted cancer therapies in the twenty-first century: lessons from imatinib.
\newblock \emph{Clin. Pharmacol. Ther.} 87:543--552.

\bibitem[Reddy and Aggarwal(2012)]{reddy2012ins}
Reddy, E.~P., and A.~K. Aggarwal, 2012.
\newblock The ins and outs of bcr-abl inhibition.
\newblock \emph{Genes \& cancer} 3:447--454.

\bibitem[Shoichet(2004)]{shoichet2004virtual}
Shoichet, B.~K., 2004.
\newblock Virtual screening of chemical libraries.
\newblock \emph{Nature} 432:862--865.

\bibitem[Okimoto et~al.(2009)Okimoto, Futatsugi, Fuji, Suenaga, Morimoto, Yanai, Ohno, Narumi, and Taiji]{okimoto2009high}
Okimoto, N., N.~Futatsugi, H.~Fuji, A.~Suenaga, G.~Morimoto, R.~Yanai, Y.~Ohno, T.~Narumi, and M.~Taiji, 2009.
\newblock High-performance drug discovery: computational screening by combining docking and molecular dynamics simulations.
\newblock \emph{PLoS Comput. Biol.} 5:e1000528.

\bibitem[Jorgensen(2009)]{jorgensen2009efficient}
Jorgensen, W.~L., 2009.
\newblock Efficient drug lead discovery and optimization.
\newblock \emph{Accounts of chemical research} 42:724--733.

\bibitem[Lin et~al.(2020)Lin, Li, and Lin]{lin2020review}
Lin, X., X.~Li, and X.~Lin, 2020.
\newblock A review on applications of computational methods in drug screening and design.
\newblock \emph{Molecules} 25:1375.

\bibitem[Dellago et~al.(1998)Dellago, Bolhuis, Csajka, and Chandler]{dellago1998transition}
Dellago, C., P.~G. Bolhuis, F.~S. Csajka, and D.~Chandler, 1998.
\newblock Transition path sampling and the calculation of rate constants.
\newblock \emph{J. Chem. Phys.} 108:1964--1977.

\bibitem[Bolhuis et~al.(2002)Bolhuis, Chandler, Dellago, and Geissler]{bolhuis2002transition}
Bolhuis, P.~G., D.~Chandler, C.~Dellago, and P.~L. Geissler, 2002.
\newblock Transition path sampling: Throwing ropes over rough mountain passes, in the dark.
\newblock \emph{Annu. Rev. Phys. Chem.} 53:291--318.

\bibitem[Bolhuis and Swenson(2021{\natexlab{a}})]{bolhuis2021transition}
Bolhuis, P.~G., and D.~W. Swenson, 2021.
\newblock Transition path sampling as Markov chain Monte Carlo of trajectories: Recent algorithms, software, applications, and future outlook.
\newblock \emph{Adv. Theory Simul.} 4:2000237.

\bibitem[Bolhuis and Swenson(2021{\natexlab{b}})]{bolhuis_transition_2021}
Bolhuis, P.~G., and D.~W.~H. Swenson, 2021.
\newblock Transition {Path} {Sampling} as {Markov} {Chain} {Monte} {Carlo} of {Trajectories}: {Recent} {Algorithms}, {Software}, {Applications}, and {Future} {Outlook}.
\newblock \emph{Adv. Theory Simul.} 4:2000237.

\bibitem[Van~Erp et~al.(2003)Van~Erp, Moroni, and Bolhuis]{van2003novel}
Van~Erp, T.~S., D.~Moroni, and P.~G. Bolhuis, 2003.
\newblock A novel path sampling method for the calculation of rate constants.
\newblock \emph{J. Chem. Phys.} 118:7762--7774.

\bibitem[Van~Erp(2012)]{van2012dynamical}
Van~Erp, T.~S., 2012.
\newblock Dynamical rare event simulation techniques for equilibrium and nonequilibrium systems.
\newblock \emph{Adv. Chem. Phys.} 151:27.

\bibitem[Faradjian and Elber(2004)]{faradjian2004computing}
Faradjian, A.~K., and R.~Elber, 2004.
\newblock Computing time scales from reaction coordinates by milestoning.
\newblock \emph{J. Chem. Phys.} 120:10880--10889.

\bibitem[Ojha et~al.(2024)Ojha, Votapka, and Amaro]{ojha2024advances}
Ojha, A.~A., L.~W. Votapka, and R.~E. Amaro, 2024.
\newblock Advances and Challenges in Milestoning Simulations for Drug–Target Kinetics.
\newblock \emph{Journal of Chemical Theory and Computation} 20:9759--9769.
\newblock PMID: 39508322.

\bibitem[Moroni et~al.(2004)Moroni, Bolhuis, and van Erp]{moroni2004rate}
Moroni, D., P.~G. Bolhuis, and T.~S. van Erp, 2004.
\newblock Rate constants for diffusive processes by partial path sampling.
\newblock \emph{J. Chem. Phys.} 120:4055--4065.

\bibitem[van Erp(2007)]{van2007reaction}
van Erp, T.~S., 2007.
\newblock Reaction rate calculation by parallel path swapping.
\newblock \emph{Phys. Rev. Lett.} 98:268301.

\bibitem[Bolhuis(2008)]{bolhuis2008rare}
Bolhuis, P.~G., 2008.
\newblock Rare events via multiple reaction channels sampled by path replica exchange.
\newblock \emph{J. Chem. Phys.} 129.

\bibitem[Cabriolu et~al.(2017)Cabriolu, Skjelbred~Refsnes, Bolhuis, and van Erp]{cabriolu2017foundations}
Cabriolu, R., K.~M. Skjelbred~Refsnes, P.~G. Bolhuis, and T.~S. van Erp, 2017.
\newblock Foundations and latest advances in replica exchange transition interface sampling.
\newblock \emph{J. Chem. Phys.} 147.

\bibitem[Vervust et~al.(2023)Vervust, Zhang, Van~Erp, and Ghysels]{vervust2023path}
Vervust, W., D.~T. Zhang, T.~S. Van~Erp, and A.~Ghysels, 2023.
\newblock Path sampling with memory reduction and replica exchange to reach long permeation timescales.
\newblock \emph{Biophys. J.} 122:2960--2972.

\bibitem[Zhang et~al.(2024)Zhang, Baldauf, Roet, Lervik, and van Erp]{zhang2024highly}
Zhang, D.~T., L.~Baldauf, S.~Roet, A.~Lervik, and T.~S. van Erp, 2024.
\newblock Highly parallelizable path sampling with minimal rejections using asynchronous replica exchange and infinite swaps.
\newblock \emph{Proc. Natl. Acad. Sci. U.S.A.} 121:e2318731121.

\bibitem[Moqadam et~al.(2017)Moqadam, Riccardi, Trinh, Lervik, and van Erp]{Mahmoud_silic}
Moqadam, M., E.~Riccardi, T.~T. Trinh, A.~Lervik, and T.~S. van Erp, 2017.
\newblock Rare event simulations reveal subtle key steps in aqueous silicate condensation.
\newblock \emph{Phys. Chem. Chem. Phys.} 19:13361--13371.

\bibitem[Moqadam et~al.(2018)Moqadam, Lervik, Riccardi, Venkatraman, Alsberg, and van Erp]{moqadamlocal2018}
Moqadam, M., A.~Lervik, E.~Riccardi, V.~Venkatraman, B.~K. Alsberg, and T.~S. van Erp, 2018.
\newblock Local initiation conditions for water autoionization.
\newblock \emph{Proc. Nat. Acad. Sci. U.S.A.} 115:E4569--E4576.

\bibitem[Daub et~al.(2020)Daub, Riccardi, H{\"a}nninen, and Halonen]{daub2020path}
Daub, C.~D., E.~Riccardi, V.~H{\"a}nninen, and L.~Halonen, 2020.
\newblock Path sampling for atmospheric reactions: formic acid catalysed conversion of SO3+ H2O to H2SO4.
\newblock \emph{PeerJ Phys. Chem.} 2:e7.

\bibitem[Aar{\o}en et~al.(2022)Aar{\o}en, Riccardi, van Erp, and Sletmoen]{aaroen2022thin}
Aar{\o}en, O., E.~Riccardi, T.~S. van Erp, and M.~Sletmoen, 2022.
\newblock Thin film breakage in oil--in--water emulsions, a multidisciplinary study.
\newblock \emph{Colloids Surf. A Physicochem. Eng. Asp.} 632:127808.

\bibitem[Zhang et~al.(2023)Zhang, Riccardi, and van Erp]{zhang2023enhanced}
Zhang, D.~T., E.~Riccardi, and T.~S. van Erp, 2023.
\newblock Enhanced path sampling using subtrajectory Monte Carlo moves.
\newblock \emph{J. Chem. Phys.} 158.

\bibitem[Lervik et~al.(2022)Lervik, Svenum, Wang, Cabriolu, Riccardi, Andersson, and van Erp]{lervik2022role}
Lervik, A., I.-H. Svenum, Z.~Wang, R.~Cabriolu, E.~Riccardi, S.~Andersson, and T.~S. van Erp, 2022.
\newblock The role of pressure and defects in the wurtzite to rock salt transition in cadmium selenide.
\newblock \emph{Phys. Chem. Chem. Phys.} 24:8378--8386.

\bibitem[Riccardi et~al.(2019)Riccardi, van Mastbergen, Navarre, and Vreede]{riccardi2019predicting}
Riccardi, E., E.~C. van Mastbergen, W.~W. Navarre, and J.~Vreede, 2019.
\newblock Predicting the mechanism and rate of {H-NS} binding to {AT-rich DNA}.
\newblock \emph{PLoS Comput. Biol.} 15:e1006845.

\bibitem[Riccardi et~al.(2020{\natexlab{a}})Riccardi, Kr{\"a}mer, van Erp, and Ghysels]{riccardi2020permeation}
Riccardi, E., A.~Kr{\"a}mer, T.~S. van Erp, and A.~Ghysels, 2020.
\newblock Permeation rates of oxygen through a lipid bilayer using replica exchange transition interface sampling.
\newblock \emph{J. Phys. Chem. B} 125:193--201.

\bibitem[Ghysels et~al.(2021)Ghysels, Roet, Davoudi, and van Erp]{Ghysels2021exactnonMarkov}
Ghysels, A., S.~Roet, S.~Davoudi, and T.~S. van Erp, 2021.
\newblock Exact non-Markovian permeability from rare event simulations.
\newblock \emph{Phys. Rev. Res.} 3:033068.

\bibitem[Panjarian et~al.(2013)Panjarian, Iacob, Chen, Engen, and Smithgall]{panjarian2013structure}
Panjarian, S., R.~E. Iacob, S.~Chen, J.~R. Engen, and T.~E. Smithgall, 2013.
\newblock Structure and dynamic regulation of Abl kinases.
\newblock \emph{J. Biol. Chem.} 288:5443--5450.

\bibitem[Simpson et~al.(2019)Simpson, Bertrand, Borthwick, Campobasso, Chabanet, Chen, Coggins, Cottom, Christensen, Dawson, et~al.]{simpson2019identification}
Simpson, G.~L., S.~M. Bertrand, J.~A. Borthwick, N.~Campobasso, J.~Chabanet, S.~Chen, J.~Coggins, J.~Cottom, S.~B. Christensen, H.~C. Dawson, et~al., 2019.
\newblock Identification and optimization of novel small c-Abl kinase activators using fragment and HTS methodologies.
\newblock \emph{J. Med. Chem.} 62:2154--2171.

\bibitem[Xie et~al.(2020)Xie, Saleh, Rossi, and Kalodimos]{xie2020conformational}
Xie, T., T.~Saleh, P.~Rossi, and C.~G. Kalodimos, 2020.
\newblock Conformational states dynamically populated by a kinase determine its function.
\newblock \emph{Science} 370:eabc2754.

\bibitem[DeLano et~al.(2002)]{delano2002pymol}
DeLano, W.~L., et~al., 2002.
\newblock Pymol: An open-source molecular graphics tool.
\newblock \emph{CCP4 Newsl. Protein Crystallogr} 40:82--92.

\bibitem[Shah et~al.(2002)Shah, Nicoll, Nagar, Gorre, Paquette, Kuriyan, and Sawyers]{shah2002multiple}
Shah, N.~P., J.~M. Nicoll, B.~Nagar, M.~E. Gorre, R.~L. Paquette, J.~Kuriyan, and C.~L. Sawyers, 2002.
\newblock Multiple BCR-ABL kinase domain mutations confer polyclonal resistance to the tyrosine kinase inhibitor imatinib (STI571) in chronic phase and blast crisis chronic myeloid leukemia.
\newblock \emph{Cancer cell} 2:117--125.

\bibitem[Shah et~al.(2007)Shah, Skaggs, Branford, Hughes, Nicoll, Paquette, Sawyers, et~al.]{shah2007sequential}
Shah, N.~P., B.~J. Skaggs, S.~Branford, T.~P. Hughes, J.~M. Nicoll, R.~L. Paquette, C.~L. Sawyers, et~al., 2007.
\newblock Sequential ABL kinase inhibitor therapy selects for compound drug-resistant BCR-ABL mutations with altered oncogenic potency.
\newblock \emph{The Journal of clinical investigation} 117:2562--2569.

\bibitem[van Erp(2023)]{van2023far}
van Erp, T.~S., 2023.
\newblock How far can we stretch the timescale with RETIS?
\newblock \emph{EPL} 143:30001.

\bibitem[Gr{\"u}nwald et~al.(2008)Gr{\"u}nwald, Dellago, and Geissler]{grunwald2008precision}
Gr{\"u}nwald, M., C.~Dellago, and P.~L. Geissler, 2008.
\newblock Precision shooting: Sampling long transition pathways.
\newblock \emph{J. Chem. Phys.} 129.

\bibitem[Gingrich and Geissler(2015)]{gingrich2015preserving}
Gingrich, T.~R., and P.~L. Geissler, 2015.
\newblock Preserving correlations between trajectories for efficient path sampling.
\newblock \emph{J. Chem. Phys.} 142.

\bibitem[Jung et~al.(2017)Jung, Okazaki, and Hummer]{jung2017transition}
Jung, H., K.-i. Okazaki, and G.~Hummer, 2017.
\newblock Transition path sampling of rare events by shooting from the top.
\newblock \emph{J. Chem. Phys.} 147.

\bibitem[Menzl et~al.(2016)Menzl, Singraber, and Dellago]{menzl2016s}
Menzl, G., A.~Singraber, and C.~Dellago, 2016.
\newblock S-shooting: a Bennett--Chandler-like method for the computation of rate constants from committor trajectories.
\newblock \emph{Faraday Discussions} 195:345--364.

\bibitem[Borrero and Dellago(2016)]{borrero2016avoiding}
Borrero, E.~E., and C.~Dellago, 2016.
\newblock Avoiding traps in trajectory space: Metadynamics enhanced transition path sampling.
\newblock \emph{Eur. Phys. J. Spec. Top.} 225:1609--1620.

\bibitem[Riccardi et~al.(2017)Riccardi, Dahlen, and van Erp]{riccardi2017fast}
Riccardi, E., O.~Dahlen, and T.~S. van Erp, 2017.
\newblock Fast Decorrelating Monte Carlo Moves for Efficient Path Sampling.
\newblock \emph{J. Phys. Chem. Lett.} 8:4456--4460.

\bibitem[Roet et~al.(2022)Roet, Zhang, and van Erp]{roet2022exchanging}
Roet, S., D.~T. Zhang, and T.~S. van Erp, 2022.
\newblock Exchanging replicas with unequal cost, infinitely and permanently.
\newblock \emph{J. Phys. Chem. A} 126:8878--8886.

\bibitem[Jo et~al.(2008)Jo, Kim, Iyer, and Im]{jo2008charmm}
Jo, S., T.~Kim, V.~G. Iyer, and W.~Im, 2008.
\newblock CHARMM-GUI: a web-based graphical user interface for CHARMM.
\newblock \emph{J. Comput. Chem.} 29:1859--1865.

\bibitem[Chan et~al.(2011)Chan, Wise, Kaufman, Ahn, Ensinger, Haack, Hood, Jones, Lord, Lu, et~al.]{chan2011conformational}
Chan, W.~W., S.~C. Wise, M.~D. Kaufman, Y.~M. Ahn, C.~L. Ensinger, T.~Haack, M.~M. Hood, J.~Jones, J.~W. Lord, W.~P. Lu, et~al., 2011.
\newblock Conformational control inhibition of the BCR-ABL1 tyrosine kinase, including the gatekeeper T315I mutant, by the switch-control inhibitor DCC-2036.
\newblock \emph{Cancer cell} 19:556--568.

\bibitem[Abraham et~al.(2015)Abraham, Murtola, Schulz, P{\'a}ll, Smith, Hess, and Lindahl]{abraham2015gromacs}
Abraham, M.~J., T.~Murtola, R.~Schulz, S.~P{\'a}ll, J.~C. Smith, B.~Hess, and E.~Lindahl, 2015.
\newblock GROMACS: High performance molecular simulations through multi-level parallelism from laptops to supercomputers.
\newblock \emph{SoftwareX} 1:19--25.

\bibitem[Huang et~al.(2017)Huang, Rauscher, Nawrocki, Ran, Feig, De~Groot, Grubm{\"u}ller, and MacKerell~Jr]{huang2017charmm36m}
Huang, J., S.~Rauscher, G.~Nawrocki, T.~Ran, M.~Feig, B.~L. De~Groot, H.~Grubm{\"u}ller, and A.~D. MacKerell~Jr, 2017.
\newblock CHARMM36m: an improved force field for folded and intrinsically disordered proteins.
\newblock \emph{Nat. Methods} 14:71--73.

\bibitem[Jorgensen et~al.(1983)Jorgensen, Chandrasekhar, Madura, Impey, and Klein]{jorgensen1983comparison}
Jorgensen, W.~L., J.~Chandrasekhar, J.~D. Madura, R.~W. Impey, and M.~L. Klein, 1983.
\newblock Comparison of simple potential functions for simulating liquid water.
\newblock \emph{J. Chem. Phys.} 79:926--935.

\bibitem[Lin et~al.(2014)Lin, Meng, Huang, and Roux]{lin2014computational}
Lin, Y.-L., Y.~Meng, L.~Huang, and B.~Roux, 2014.
\newblock Computational study of Gleevec and G6G reveals molecular determinants of kinase inhibitor selectivity.
\newblock \emph{J. Am. Chem. Soc.} 136:14753--14762.

\bibitem[Huang and Roux(2013)]{huang2013automated}
Huang, L., and B.~Roux, 2013.
\newblock Automated force field parameterization for nonpolarizable and polarizable atomic models based on ab initio target data.
\newblock \emph{J. Chem. Theory Comput.} 9:3543--3556.

\bibitem[Boulanger et~al.(2018)Boulanger, Huang, Rupakheti, MacKerell~Jr, and Roux]{boulanger2018optimized}
Boulanger, E., L.~Huang, C.~Rupakheti, A.~D. MacKerell~Jr, and B.~Roux, 2018.
\newblock Optimized Lennard-Jones parameters for druglike small molecules.
\newblock \emph{J. Chem. Theory Comput.} 14:3121--3131.

\bibitem[Paul et~al.(2020)Paul, Thomas, and Roux]{paul2020diversity}
Paul, F., T.~Thomas, and B.~Roux, 2020.
\newblock Diversity of long-lived intermediates along the binding pathway of imatinib to Abl kinase revealed by MD simulations.
\newblock \emph{J. Chem. Theory Comput.} 16:7852--7865.

\bibitem[Aleksandrov and Simonson(2010)]{aleksandrov2010molecular}
Aleksandrov, A., and T.~Simonson, 2010.
\newblock A molecular mechanics model for imatinib and imatinib: kinase binding.
\newblock \emph{J. Comput. Chem.} 31:1550--1560.

\bibitem[Bussi et~al.(2007)Bussi, Donadio, and Parrinello]{bussi2007canonical}
Bussi, G., D.~Donadio, and M.~Parrinello, 2007.
\newblock Canonical sampling through velocity rescaling.
\newblock \emph{J. Chem. Phys.} 126.

\bibitem[Parrinello and Rahman(1981)]{parrinello1981polymorphic}
Parrinello, M., and A.~Rahman, 1981.
\newblock Polymorphic transitions in single crystals: A new molecular dynamics method.
\newblock \emph{J. Appl. Phys.} 52:7182--7190.

\bibitem[Hess et~al.(1997)Hess, Bekker, Berendsen, and Fraaije]{hess1997lincs}
Hess, B., H.~Bekker, H.~J. Berendsen, and J.~G. Fraaije, 1997.
\newblock LINCS: A linear constraint solver for molecular simulations.
\newblock \emph{J. Comput. Chem.} 18:1463--1472.

\bibitem[Humphrey et~al.(1996)Humphrey, Dalke, and Schulten]{humphrey1996vmd}
Humphrey, W., A.~Dalke, and K.~Schulten, 1996.
\newblock VMD: visual molecular dynamics.
\newblock \emph{J. Mol. Graph.} 14:33--38.

\bibitem[Tribello et~al.(2014)Tribello, Bonomi, Branduardi, Camilloni, and Bussi]{tribello2014plumed}
Tribello, G.~A., M.~Bonomi, D.~Branduardi, C.~Camilloni, and G.~Bussi, 2014.
\newblock PLUMED 2: New feathers for an old bird.
\newblock \emph{Comput. Phys. Commun.} 185:604--613.

\bibitem[Riccardi et~al.(2020{\natexlab{b}})Riccardi, Lervik, Roet, Aar{\o}en, and van Erp]{riccardi2020pyretis}
Riccardi, E., A.~Lervik, S.~Roet, O.~Aar{\o}en, and T.~S. van Erp, 2020.
\newblock PyRETIS 2: an improbability drive for rare events.
\newblock \emph{Journal of computational chemistry} 41:370--377.

\bibitem[inf()]{infretissoftware}
InfRETIS: Python libraries for running $\infty$RETIS.
\newblock \urlprefix\url{https://github.com/infretis/}, gitHub repository.

\bibitem[Vervust et~al.(2024)Vervust, Zhang, Ghysels, Roet, van Erp, and Riccardi]{vervust2024pyretis}
Vervust, W., D.~T. Zhang, A.~Ghysels, S.~Roet, T.~S. van Erp, and E.~Riccardi, 2024.
\newblock PyRETIS 3: Conquering rare and slow events without boundaries.
\newblock \emph{J. Comput. Chem.} .

\bibitem[Seeliger et~al.(2007)Seeliger, Nagar, Frank, Cao, Henderson, and Kuriyan]{seeliger2007c}
Seeliger, M.~A., B.~Nagar, F.~Frank, X.~Cao, M.~N. Henderson, and J.~Kuriyan, 2007.
\newblock c-Src binds to the cancer drug imatinib with an inactive Abl/c-Kit conformation and a distributed thermodynamic penalty.
\newblock \emph{Structure} 15:299--311.

\bibitem[Lyczek et~al.(2021)Lyczek, Berger, Rangwala, Paung, Tom, Philipose, Guo, Albanese, Robers, Knapp, Chodera, and Seeliger]{lyczek2021mutation}
Lyczek, A., B.-T. Berger, A.~M. Rangwala, Y.~Paung, J.~Tom, H.~Philipose, J.~Guo, S.~K. Albanese, M.~B. Robers, S.~Knapp, J.~D. Chodera, and M.~A. Seeliger, 2021.
\newblock Mutation in Abl kinase with altered drug-binding kinetics indicates a novel mechanism of imatinib resistance.
\newblock \emph{Proc. Natl. Acad. Sci. U.S.A.} 118:e2111451118.

\bibitem[Agafonov et~al.(2014)Agafonov, Wilson, Otten, Buosi, and Kern]{agafonov2014energetic}
Agafonov, R.~V., C.~Wilson, R.~Otten, V.~Buosi, and D.~Kern, 2014.
\newblock Energetic dissection of Gleevec's selectivity toward human tyrosine kinases.
\newblock \emph{Nat. Struct. Mol. Biol.} 21:848--853.

\bibitem[Narayan et~al.(2021)Narayan, Buchete, and Elber]{narayan2021computer}
Narayan, B., N.-V. Buchete, and R.~Elber, 2021.
\newblock Computer simulations of the dissociation mechanism of Gleevec from Abl Kinase with milestoning.
\newblock \emph{J. Phys. Chem. B} 125:5706--5715.

\bibitem[Shekhar et~al.(2022)Shekhar, Smith, Seeliger, and Tiwary]{shekhar2022protein}
Shekhar, M., Z.~Smith, M.~A. Seeliger, and P.~Tiwary, 2022.
\newblock Protein flexibility and dissociation pathway differentiation can explain onset of resistance mutations in kinases.
\newblock \emph{Angew. Chem. Int. Ed.} 61:e202200983.

\bibitem[van Erp(2006)]{van2006efficiency}
van Erp, T.~S., 2006.
\newblock Efficiency analysis of reaction rate calculation methods using analytical models I: The two-dimensional sharp barrier.
\newblock \emph{J. Chem. Phys.} 125.

\bibitem[Allen et~al.(2009)Allen, Valeriani, and Ten~Wolde]{allen2009forward}
Allen, R.~J., C.~Valeriani, and P.~R. Ten~Wolde, 2009.
\newblock Forward flux sampling for rare event simulations.
\newblock \emph{J. Phys.: Condens. Matter} 21:463102.

\bibitem[Frenkel and Smit(2023)]{frenkel2023understanding}
Frenkel, D., and B.~Smit, 2023.
\newblock Understanding molecular simulation: from algorithms to applications.
\newblock Elsevier.

\bibitem[Huber and Kim(1996)]{huber1996weighted}
Huber, G.~A., and S.~Kim, 1996.
\newblock Weighted-ensemble Brownian dynamics simulations for protein association reactions.
\newblock \emph{Biophysical journal} 70:97--110.

\bibitem[C{\'e}rou et~al.(2011)C{\'e}rou, Guyader, Lelievre, and Pommier]{cerou2011multiple}
C{\'e}rou, F., A.~Guyader, T.~Lelievre, and D.~Pommier, 2011.
\newblock A multiple replica approach to simulate reactive trajectories.
\newblock \emph{J. Chem. Phys.} 134.

\bibitem[van Heesch et~al.(2023)van Heesch, Bolhuis, and Vreede]{van2023decoding}
van Heesch, T., P.~G. Bolhuis, and J.~Vreede, 2023.
\newblock Decoding dissociation of sequence-specific protein--DNA complexes with non-equilibrium simulations.
\newblock \emph{Nucleic Acids Research} 51:12150--12160.

\bibitem[Weinan et~al.(2002)Weinan, Ren, and Vanden-Eijnden]{weinan2002string}
Weinan, E., W.~Ren, and E.~Vanden-Eijnden, 2002.
\newblock String method for the study of rare events.
\newblock \emph{Physical Review B} 66:052301.

\bibitem[Maragliano et~al.(2006)Maragliano, Fischer, Vanden-Eijnden, and Ciccotti]{maragliano2006string}
Maragliano, L., A.~Fischer, E.~Vanden-Eijnden, and G.~Ciccotti, 2006.
\newblock String method in collective variables: Minimum free energy paths and isocommittor surfaces.
\newblock \emph{The Journal of chemical physics} 125.

\bibitem[Frassek et~al.(2021)Frassek, Arjun, and Bolhuis]{frassek2021extended}
Frassek, M., A.~Arjun, and P.~Bolhuis, 2021.
\newblock An extended autoencoder model for reaction coordinate discovery in rare event molecular dynamics datasets.
\newblock \emph{The Journal of Chemical Physics} 155.

\bibitem[Belkacemi et~al.(2021)Belkacemi, Gkeka, Leli{\`e}vre, and Stoltz]{belkacemi2021chasing}
Belkacemi, Z., P.~Gkeka, T.~Leli{\`e}vre, and G.~Stoltz, 2021.
\newblock Chasing collective variables using autoencoders and biased trajectories.
\newblock \emph{Journal of chemical theory and computation} 18:59--78.

\bibitem[Fr{\"o}hlking et~al.(2024)Fr{\"o}hlking, Bonati, Rizzi, and Gervasio]{frohlking2024deep}
Fr{\"o}hlking, T., L.~Bonati, V.~Rizzi, and F.~L. Gervasio, 2024.
\newblock Deep learning path-like collective variable for enhanced sampling molecular dynamics.
\newblock \emph{The Journal of Chemical Physics} 160.

\bibitem[Ayaz et~al.(2023)Ayaz, Lyczek, Paung, Mingione, Iacob, de~Waal, Engen, Seeliger, Shan, and Shaw]{ayaz2023structural}
Ayaz, P., A.~Lyczek, Y.~Paung, V.~R. Mingione, R.~E. Iacob, P.~W. de~Waal, J.~R. Engen, M.~A. Seeliger, Y.~Shan, and D.~E. Shaw, 2023.
\newblock Structural mechanism of a drug-binding process involving a large conformational change of the protein target.
\newblock \emph{Nat. Commun.} 14:1885.

\bibitem[Rogal and Bolhuis(2008)]{rogal2008multiple}
Rogal, J., and P.~G. Bolhuis, 2008.
\newblock Multiple state transition path sampling.
\newblock \emph{J. Chem. Phys.} 129.

\bibitem[Wu and No{\'e}(2020)]{wu2020variational}
Wu, H., and F.~No{\'e}, 2020.
\newblock Variational approach for learning Markov processes from time series data.
\newblock \emph{Journal of Nonlinear Science} 30:23--66.

\bibitem[Vervust(2024)]{vervust2024Thesis}
Vervust, W., 2024.
\newblock Advanced Molecular Dynamics Simulation Techniques for Kinetic Analysis of Biological Systems.
\newblock Doctoral dissertation, Ghent University, Faculty of Engineering and Architecture, Department of Electronics and Information Systems.

\bibitem[Falkner et~al.(2024)Falkner, Coretti, and Dellago]{Falkner2024Enhanced}
Falkner, S., A.~Coretti, and C.~Dellago, 2024.
\newblock Enhanced Sampling of Configuration and Path Space in a Generalized Ensemble by Shooting Point Exchange.
\newblock \emph{Phys. Rev. Lett.} 132:128001.

\bibitem[Vervust et~al.()Vervust, Zhang, Riccardi, {van Erp}, and Ghysels]{zeenoodoo_abl}
Vervust, W., T.~D. Zhang, E.~Riccardi, T.~S. {van Erp}, and A.~Ghysels.
\newblock ABL configurations.
\newblock Dataset on Zenodo.
\newblock \urlprefix\url{https://doi.org/10.5281/zenodo.14833469}.

\end{thebibliography}

\end{document}